\documentclass{uwstat518}

\usepackage[utf8]{inputenc} 
\usepackage[T1]{fontenc}    
\usepackage{hyperref}       
\usepackage{url}            
\usepackage{booktabs}       
\usepackage{amsfonts}       
\usepackage{nicefrac}       
\usepackage{microtype}      

\usepackage{graphicx}
\usepackage{amssymb}
\usepackage{epstopdf}
\usepackage{mathtools}
\usepackage{amsmath}
\usepackage{algorithm2e}
\usepackage{setspace}
\usepackage{booktabs}
\usepackage{array}
\newcommand{\Pj}{{\bf P}^j}
\newcommand{\Zj}{{\bf Z}^j}
\newcommand{\pzero}{{\bf p}^0}
\newcommand{\zzero}{{\bf z}^0}

\newcommand{\bone}{{\bf 1}}

\newtheorem{ntheorem}{Theorem}

\newtheorem{proposition}[ntheorem]{Proposition}

\newtheorem{definition}[ntheorem]{Definition}

\usepackage{color}

\title{Granger Causality Networks for Categorical Time Series}

\begin{document}
\RestyleAlgo{boxruled}

\title{Granger Causality Networks for Categorical Time Series}

\author{
	Alex Tank \\
       Department of Statistics \\
       University of Washington \\
       alextank@uw.edu
	\and 
	Emily Fox \\
       Department of Statistics \\
       University of Washington \\
       ebfox@uw.edu
       \and
       Ali Shojaie \\
Department of Biostatistics\\
University of Washington\\
ashojaie@uw.edu
}


\maketitle

\begin{abstract}
We present a new framework for learning Granger causality  networks for multivariate categorical time series, based on the mixture transition distribution (MTD) model. Traditionally, MTD is plagued by a nonconvex objective, non-identifiability, and presence of many local optima. To circumvent these problems, we recast inference in the MTD as a convex problem. The new formulation facilitates the application of MTD to high-dimensional multivariate time series. 
As a baseline, we also formulate a multi-output logistic autoregressive model (mLTD), which while a straightforward extension of autoregressive Bernoulli generalized linear models, has not been previously applied to the analysis of multivariate categorial time series. 
We develop novel identifiability conditions of the MTD model and compare them to those for mLTD. We further devise novel and efficient optimization algorithm for the MTD based on the new convex formulation, and compare the MTD and mLTD in both simulated and real data experiments. Our approach simultaneously provides a comparison of methods for network inference in categorical time series and opens the door to modern, regularized inference with the MTD model.

\end{abstract}

\section{Introduction}
Granger causality~\cite{Granger:1980} is a popular framework for assessing the relationships between time series, and has been widely applied in econometrics, neuroscience, and genomics, amongst other fields. Given two time series $x$ and $y$, the idea is to use the temporal structure of the data to assess whether the past values of one, say $x$, are predictive of future values of the other, $y$, beyond what the past of $y$ can predict alone; if so, $x$ is said to \emph{Granger cause} $y$.  Recently, the focus has shifted to inferring Granger causality networks from multivariate time series data, with the goal of uncovering a sparse set of Granger causal relationships amongst the individual univariate time series.  Building on the typical autoregressive framework for assessing Granger causality, a majority of approaches for inferring Granger causal networks have focused on real-valued Gaussian time series using the vector autoregressive model (VAR) with sparsity inducing penalties \cite{Han:2013,Shojaie:2010}. More recently, this approach has been extended to non-Gaussian data such as multivariate point processes using sparse Hawkes processes \cite{Zhou:2013}, count data using autoregressive Poisson generalized linear models \cite{Hall:2016}, or even time series with heavy tails using VAR models with elliptical errors \cite{qiu:2015}. In contrast, inferring networks for multivariate \emph{categorical} time series has not been studied under this paradigm.

Multivariate categorical time series arise naturally in many domains. For example, we might have health states from various indicators for a patient over time, voting records for a set of politicians, action labels for players on a team, social behaviors for kids in a school, or musical notes in an orchestrated piece.  There are also many datasets that can be viewed as binary multivariate time series based on the presence or absence of an action for some set of entities.  Furthermore, in some applications, collections of continuous-valued time series are each quantized into a set of discrete values, like the weather data from multiple stations analyzed in \cite{doshi:2011}, wind data in \cite{Raftery:1985}, stock returns in \cite{Nicolau:2014}, or sales volume for a collection of products in \cite{Ching:2002}.

The \emph{mixture transition distribution} (MTD) model \cite{berchtold:2002,Raftery:1985}, originally proposed for parsimonious modelling of higher order Markov chains, can provide an approach to modeling multivariate categorical time series \cite{Ching:2002,Nicolau:2014,Zhu:2010}. The MTD model reduces each categorical interaction to a standard single dimensional Markov transition probability table. While alluring due to its elegant construction and intuitive interpretation, widespread use of the MTD model has been limited by a non-convex objective with many local optima, a large number of parameter constraints, and unknown identifiability conditions \cite{Nicolau:2014,Zhu:2010,Berchtold:2001}. For this reason, most applications of the MTD model to multivariate time series have looked at a maximum of three or four time series. To bypass the limitations of MTD, autoregressive generalized linear models have been advocated for categorical time series. In particular, autoregressive generalized linear binomial models are often used for the special case of two categories per series \cite{Hall:2016, bahadori:2013}. However, their multinomial-output extension to a larger number of states per series has not been widely adopted. See \cite{kedem:2005} for an application to the univariate time series case. 

We refer to the autoregressive multinomial GLM as the mixture logistic transition distribution (mLTD). The mLTD model uses a logistic function to bypass parameter constraints, results in a convex objective, and has well-known identifiability conditions. However, these advantages of mLTD come at the cost of reduced interpretability, mainly because the transition distribution in mLTD depends nonlinearly on the model parameters. 
\cite{Nicolau:2014} has recently proposed a constrained autoregressive probit model that improves interpretability. However, the probit model is both highly non-convex and inference is computationally intensive, limiting applications to higher dimensional series. As such, one is still torn between a computational and interpretability tradeoff. We address this issue by going back to the interpredability of the MTD framework and showing how one can dramatically improve its computationational drawbacks. 

In particular, we recast inference in the MTD model as a convex problem through a novel re-parameterization. We further develop a regularized estimation framework for identifying Granger causality for multivariate categorical time series. We also establish for the first time conditions for identifiability in the MTD model and compare the identifiability conditions for MTD and mLTD models. We find that while the identifiability conditions for the MTD model are given by a non-convex set, we may easily enforce the constraints using our convex re-parameterization trick by augmenting the likelihood with appropriate convex penalties. We then develop an efficient projected gradient algorithm for optimizing the penalized convex MTD objective. Our efficient algorithm depends on a Dykstra splitting method for projection onto the constraint sets of the MTD model. This computational approach for MTD provides enormous gains over past methods, enabling this model to be applied to large, modern datasets for the first time.  Importantly, the computational insights provided in this paper carry over to the suite of other applications of MTD models, such as higher order Markov chains, beyond the multivariate categorical time series which are the focus herein.

As a comparison benchmark we also develop a penalized mLTD model for Granger causality in multivariate Markov chains. While straightforward, the application of the penalized mLTD framework to multivariate categorical time series with more than two categories is new. We compare MTD and mLTD methods under multiple simulation conditions and use the MTD method to uncover Granger causality structure in a music data set. Studying the potential theoretical benefits of one framework over the other is left as future work.

\section{Categorical Time Series and Granger Causality}
\subsection{Granger Causality}
Let $x_t = (x_{1t}, \ldots x_{dt}), \in \mathcal{X} $ denote a $d$-dimensional categorical random variable indexed by time where $\mathcal{X} = \left(\mathcal{X}_1 \times \mathcal{X}_2 \ldots \times \mathcal{X}_d \right)$, with $\mathcal{X}_i$ denoting the set of possible values of $x_{it}$. Let $m_i = |\mathcal{X}_i|$ be the cardinality of set $\mathcal{X}_i$, i.e. the number of categories series $i$ may take. A length $T$ multivariate categorical time series is the sequence $X = \{x_1,\ldots , x_{t}, \ldots,x_T\}$. An order $k$ multivariate Markov chain models the transition probability between the categories at lagged times $t-1,\ldots, t-k$ and those at time $t$ using a transition probability tensor:
\begin{align} 
p(x_t|x_{t-1},\ldots) = p(x_t  |x_{t-1},\ldots,x_{t-k}).
\end{align}
Due to the complexity of fully parameterizing this transition distribution, it is common to simplify the model and assume that the categories at time $t$ are conditionally independent of one another given the past realizations:
\begin{align}
p(x_t  |x_{t-1},\ldots,x_{t-k}) = \prod_{i = 1}^{d} p(x_{it}  |x_{t-1},\ldots,x_{t-k}).
\label{eq:margcondind}
\end{align}
For simplicity, we assume $k=1$, but stress that all models and results equally apply to higher orders of $k$. Based on the decomposition assumption, Eq. (\ref{eq:margcondind}), the problem of estimation and inference decomposes into independent subproblems over each series $i$. Using this decomposition, we define Granger non-causality for two categorical time series $x_{it}$ and $x_{jt}$ as follows.
\begin{definition} \label{grangerdef}
Time series $x_j$ is not Granger causal for time series $x_i$ iff
\begin{align*}
p(x_{it} | &x_{1(t-1)},\ldots, x_{j(t-1)},\ldots x_{d(t-1)}) = p(x_{it} | x_{1(t-1)},\ldots, x_{(j-1)(t-1)},x_{(j+1)(t-1)},\ldots, x_{d(t-1)}).
\end{align*}
\end{definition}
Definition \ref{grangerdef} states that $x_{jt}$ is not Granger causal for time series $x_{it}$ if the probability that $x_{it}$ is in a given state at time $t$ is conditionally independent of the value of $x_{j(t-1)}$ at time $t-1$ given the values of all other series $x_{k(t-1)}$, $k\neq i,j$, at time lag $t-1$. Definition \ref{grangerdef} is natural since it implies that if $x_{it}$ does not Granger cause $x_{jt}$, then knowing $x_{i(t-1)}$ does not help predicting the future state of series $j$, $x_{jt}$. For real-valued data, classical definitions of Granger non-causality generally state that the conditional mean, in homoskedastic models, or conditional variance, in heteroskedastic models, of $x_{jt}$ do not depend on the past values $x_{it}$. Thus, Definition \ref{grangerdef} is a generalization of the classical case to multivariate categorical data, where notions like conditional mean and variance are less applicable. While this definition of Granger causality is intuitive and similar to other definitions for real-valued data, it has not been explicitly stated for multivariate categorical time series and represents a contribution of our work. 

\subsection{Tensor Representation for Categorical Time Series}

\begin{figure}
\centering
\includegraphics[width=.7\textwidth]{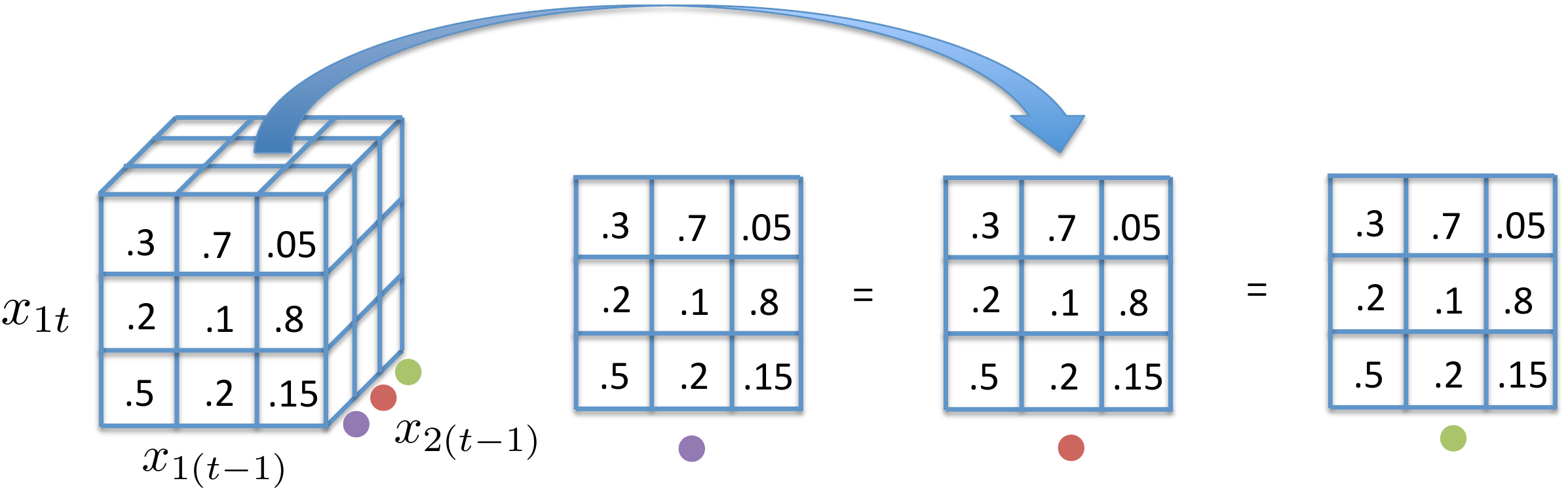}
\caption{Illustration of Granger non-causality in an example with $d = 2$ and $m_1 = m_2 = 3$. Since the tensor represents conditional probabilities, the columns of the front face of the tensor, the vertical $x_{1t}$ axis, must sum to one. Here, $x_2$ is not Granger causal for $x_1$ since each slice of the conditional probability tensor along the $x_2$ mode is equal.}
\label{tensfig}
\end{figure}

Each individual conditional distribution in Eq. (\refeq{eq:margcondind}) can be represented as a conditional probability tensor ${\bf \tilde{P}}^i$ with $p + 1$ modes of dimension $m_i \times m_1 \times \ldots \times m_d$. Each entry of the tensor is given by 
\begin{align}
{\bf \tilde{P}}^i_{x_{it},x_{1 (t-1)}, \ldots, x_{d (t-1)}} = p(x_{it} | x_{1(t-1)},\ldots, x_{j(t-1)},\ldots x_{d(t-1)}).
\end{align} 
Definition \ref{grangerdef} may be stated equivalently using the language of tensors: $x_j$ does not Granger cause $x_i$ if all unfoldings of the ${\bf \tilde{P}}^i$ tensor along the mode associated with $x_j$ are equal. This is displayed graphically in Figure \ref{tensfig}. 

The tensor interpretation suggests a naive penalized likelihood method to select for Granger non-causality in categorical time series: perform penalized maximum likelihood estimation of the conditional probability tensor with a penalty that enforces equality among the unfoldings of each mode. While we have explored the above approach in low dimensions,  $d \leq 5$, memory, and in turn, computational requirements for storing the complete probability tensor becomes infeasible for even moderate dimensions since $ {\bf \tilde{P}}^i$ has $m_i \times m_1 \times \ldots m_d$ entries. Instead, in Sections~\ref{sec:MTD} and \ref{sec:mLTD}, we present tensor parameterizations where the number of parameters needed to represent the full conditional probability tensor grows linearly with $d$.  We establish Granger non-causality conditions and associated penalized likelihood methods for estimation under these parameterizations in Sections~\ref{ident} and \ref{opt}, respectively. 

In specifying our models, and throughout the remainder of the paper, we focus in on a single conditional of $x_{it}$ given $x_{t-1}$ in Eq. (\ref{eq:margcondind}).  For notational simplicity, we drop the $i$ index; otherwise, 
\subsection{The MTD model}
\label{sec:MTD}
The MTD model \cite{Raftery:1985} provides an elegant and intuitive parameterization of the multivariate Markov transition distribution as a convex combination of pairwise transition probabilities. Specifically, the MTD model is given by:
\begin{align}
p(x_{it} | x_{1(t-1)},\ldots, &x_{d(t-1)}) = \gamma_0 p_0(x_{it}) + \sum_{j = 1}^{d} \gamma_j p_j(x_{it}|x_{j(t-1)}),
\label{eq:MTD}
\end{align}
where $p_0$ is a probability vector, $p_j(.|.)$ is a pairwise transition probability table between $x_{j(t-1)}$ and $x_{it}$ and $\gamma = (\gamma_0,\gamma_1, \ldots, \gamma_d)$ is a $d + 1$ dimensional probability distribution such that $\bone^T \gamma = 1$ with $\gamma_j \geq 0$, $j=0,\ldots,d$. We let the matrix $\Pj \in \mathbb{R}^{m_i \times m_j}$. Thus, $\bone^T \Pj = \bone^T$, $\Pj_{lk} \geq 0$, $l=1,\ldots,m_i$, $k=1,\ldots,m_j$. Denote the pairwise transitions ${\bf P}^j_{x_{it}, x_{j(t-1)}} = p_j(x_{it}|x_{j(t-1)})$. We also let $\pzero \in \mathbb{R}^{m_i}$ denote the intercept, where ${\bf p}^0_{x_{it}} = p_j(x_{it}|x_{j(t-1)})$. While past formulations of the MTD model neglect the $p_0$ intercept term, we show below that the intercept is crucial for model identifiability and, consequently, Granger causality inference. Finally, we note that the MTD model may be extended by adding in interaction terms for pairwise effects \cite{berchtold:2002}, such as $p_{jk}(x_{it}|x_{j(t-1)},x_{k(t-1)})$, though we focus our presentation on the simple case above.

\subsection{The mLTD model}
\label{sec:mLTD}
The multinomial logistic transition distribution (mLTD) model is given by:
\begin{align} \label{GLM}
p(x_{it}|x_{1(t-1)},&\ldots, x_{d(t-1)}) =  \frac{\exp\left(\zzero_{x_{it}} + \sum_{j = 1}^{d} \Zj_{x_{it}, x_{j(t-1)}}\right)}{\sum_{x' \in \mathcal{X}_i} \exp\left(\zzero_{x'} + \sum_{j = 1}^{d} \Zj_{x', x_{j(t-1)}}\right)}
\end{align}
where $\Zj \in \mathbb{R}^{m_i \times m_j}$ and $\zzero \in \mathbb{R}^{m_i}$. While not used before to model multivariate categorical time series with $m > 2$ categories, its close cousin, the probit model, has been utilized for this purpose \cite{Nicolau:2014}. The model in \cite{Nicolau:2014} is not a natural fit for inferring Granger causality networks both due to the non-convexity of the probit model and the non-convex constraints imposed on the $\Zj$ matrices. Note that, like the MTD model, the mLTD model naturally allows adding interaction terms, though we focus again our presentation on the simple case above.

\subsection{Comparing MTD and mLTD models}
\begin{figure}
\centering
\includegraphics[width = .6\textwidth]{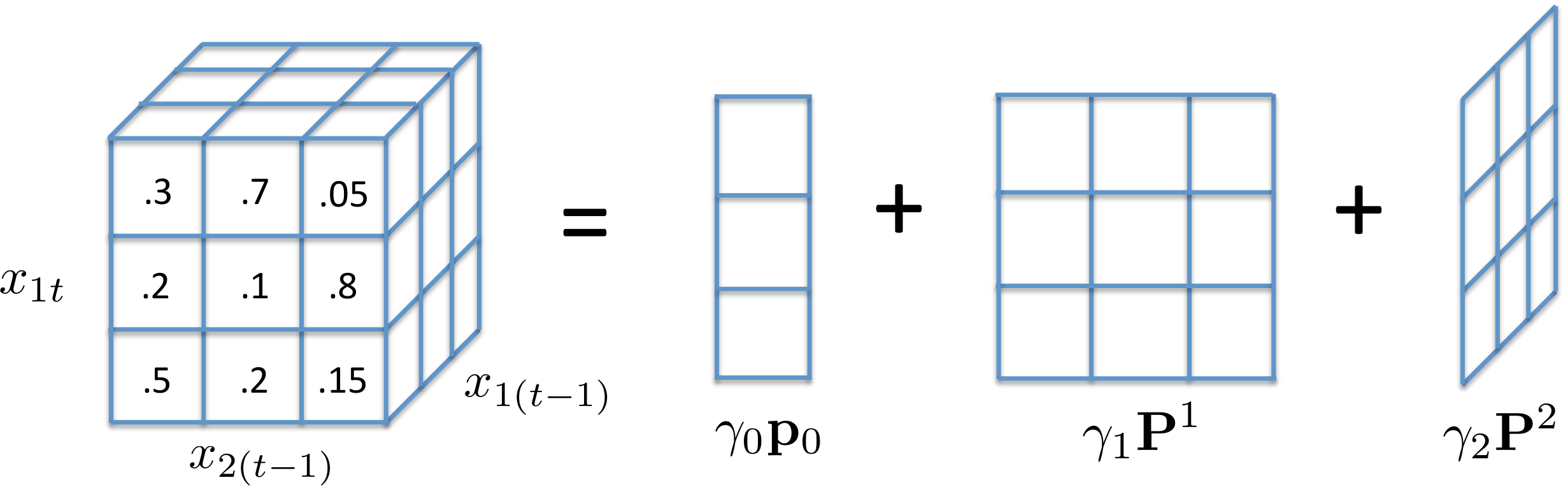}
\caption{Schematic of the MTD factorization of the conditional probability tensor $p(x_{t1} | x_{(t-1)1}, x_{(t - 1)2})$ for $d = 2$ time series and $m = 3$ categories.} \label{mtdfig}
\end{figure}

Both MTD and mLTD models represent the full conditional probability tensor using individual matrices for each $x_j$ series, $\Pj$ for MTD and $\Zj$ for mLTD. However, how these matrices are composed and restrictions on their domains differ substantially between the two models.  The MTD model is a convex combination of pairwise probability tables whereas mLTD is a nonlinear function of the unresricted $\Zj$s. MTD may thus be thought of as a linear tensor factorization method for conditional probability tensors, where the tensor is created by summing probability table slices along each dimension. This interpretation of MTD is displayed graphically in Figure \ref{mtdfig}.  

\section{Convexity, Identifiability and Granger Causality} \label{ident}

In this section, we first introduce a novel reparamaterization of the MTD model that renders the log-likelihood of the MTD model \emph{convex}.  The convex formulation alone opens up an array of possibilities for the MTD framework beyond our multivariate categorical time series focus, eliminating the primary barrier to adoption of this method, i.e. non-convexity and associated computationally demanding inference procedures. The proposed change-of-variables also allows us to derive both novel identifiability conditions for the MTD model and Granger causality restrictions that hold for both MTD and mLTD models. The non-identifiability of the MTD model was first pointed out by \cite{Lebre:2008}, but no explicit conditions or general framework for identifiability were given. We show that while the identifiability conditions for MTD are non-convex, they may be enforced implicitly by adding an appropraite convex penalty to the convex log-likelihood objective. The proofs of all results are given in the online Supplementary Material. 

\subsection{Convex MTD}

Maximum likelihood for the MTD model under the $(\gamma, {\bf P})$ parameterization is given by the non-convex optimization problem:
\begin{equation*}
\begin{aligned}
& \underset{{\bf P}, {\bf \gamma}}{\text{minimize}}   - \sum_{t = 1}^{T}\log \left(\gamma_0 \pzero_{x_{it}} + \sum_{j = 1}^{p}\gamma_{j} \Pj_{x_{it} \,\, x_{j(t-1)}} \right) \\
& \text{subject to} \,\,\,\,\, {\bf 1}^T \Pj = {\bf 1}^T ,\,\,\, \Pj \geq 0, \,\, \forall j \,\,\,\,\,\,\,\,\,\,\,\, {\bf 1}^T \gamma = 1 \,\,,\gamma \geq 0.
\end{aligned}
\end{equation*}
 The log-likelihood surface is highly non-convex, following from the multiplication of the $\gamma_j$ and $\Pj$ terms in the log term. It also contains many local optima due to the general non-identifiability. Indeed, the set of equivalent models forms a non-convex region in the $(\gamma, {\bf P})$ parameterization (i.e., the convex combination of equivalent models is not necessarily another equivalent model), leading to many non-convex shaped ridges and sets of equal probability. 

 Fortunately, optimization may be recast into a convex program using the re-parameterization $\Zj = \gamma_j \Pj$ and $\zzero = \gamma_0 \pzero $. Using this reparameterization we can rewrite the factorization of the conditional probability tensor for MTD in Eq. (\ref{eq:MTD}) as
 \begin{align} \label{MTDreparam}
 p(x_{it}|x_{1(t-1)},\ldots, x_{p(t-1)}) = \zzero_{x_{it}} + \sum_{j = 1}^{p} \Zj_{x_{it}, x_{j(t-1)}}.
 \end{align}
 The full optimization problem for maximum log-likelihood including constraints then becomes:
\begin{equation}
\begin{aligned}
& \underset{{\bf {\bf Z}}, {\bf \gamma}}{\text{minimize}}   - \sum_{t = 1}^{T}\log \left(\zzero_{x_{it}} + \sum_{j = 1}^{p} \Zj_{x_{it} \,\, x_{j(t-1)}} \right)  \\
& \text{subject to} \,\,\,\,\, {\bf 1}^T \Zj = \gamma_j{\bf 1}^T ,\,\,\, \Zj \geq 0, \,\, \forall j \,\,\,\,\,\,\,\,\,\,\,\, {\bf 1}^T \gamma = 1 \,\,,\gamma \geq 0. 
\end{aligned} 
\label{convexproblem}
\end{equation}
Problem (\ref{convexproblem}) is convex since the objective function is a linear function composed with a log function and only involves linear equality and inequality constraints \cite{boyd:2004}. 

The $\Zj$ reparameterization in Eq. (\ref{MTDreparam}) also provides clear intuition for why the MTD model may not be identifiable. Since the probability function is a linear sum of $\Zj$s, one may move probability mass around, taking some from some $\Zj$ and moving to some ${\bf Z}_i$, $i \neq j$, while keeping the conditional probability tensor constant. These sets of equivalent MTD parameterizations have the following appealing property:
\begin{proposition} \label{Convex} 
The set of MTD parameters, ${\bf Z}$, that yield the same factorized conditional distribution $p(x_{it} | x_{(t-1)})$ forms a convex set.
\end{proposition}
Taken together, the convex reparameterization and Proposition \ref{Convex} imply that the convex function given in Eq. (\ref{convexproblem}) has no local optima, and that the globaly optimal solution to Problem (\ref{convexproblem}) is given by a convex set of equivalent MTD models.

\subsection{Identifiability}

%
\subsubsection{Identifiability for the MTD model}
\label{mtd_ident_s}
\begin{figure}
\centering
\includegraphics[width=.5\textwidth]{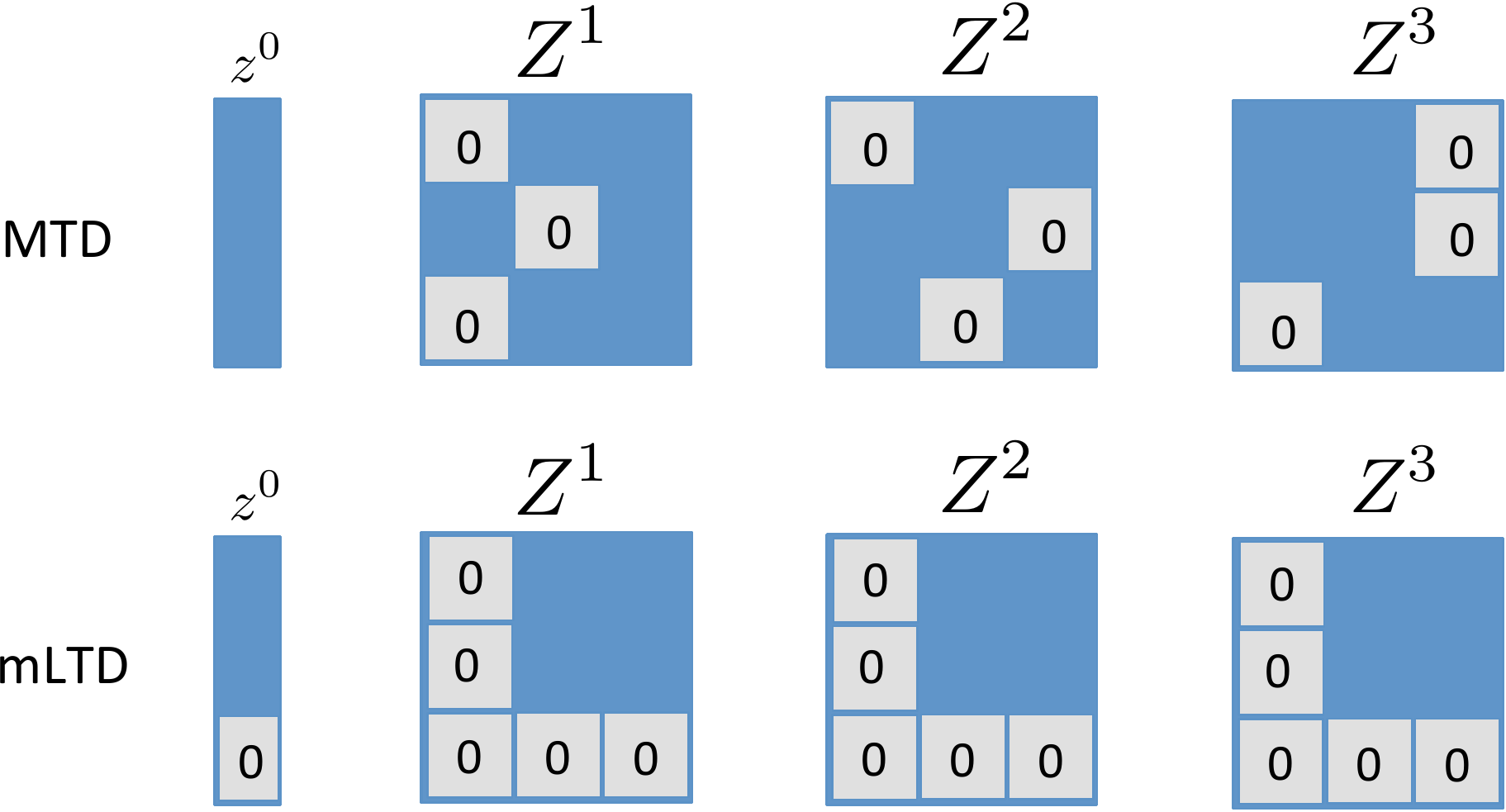}
\caption{Schematic displaying the identifiability conditions for the MTD model (\emph{top}) and the mLTD model (\emph{bottom}) for a $d = 3$ and $m_1 = m_2 = m_3 = 3$ example. Identifiability for MTD requires a zero entry in each row of $\Zj$, while for mLTD the first column and last row must all be zero. In MTD the columns of each $\Zj$ must also sum to the same value, and must sum to one across all $\Zj$.}
\label{ident_fig}
\end{figure}

The re-parameterization of the MTD model in terms of $\Zj$ instead of $\gamma_j$ and $\Pj$, combined with the introduction of an intercept term, allows us to explicitly characterize identifiability conditions for this model. 
\begin{ntheorem} \label{mtd_ident}
Every MTD distribution has a unique parameterization where the minimal element in each row of $\Pj$ (and thus $\Zj$) is zero for all $j$.
\label{identMTD}
\end{ntheorem}
The intuition for this result is simple --- any excess probability mass on a row of each $\Zj$ may be pushed onto the same row of the intercept term ${\bf z}^0$ without changing the full conditional probability. This operation may be done until the smallest element in each row is zero, but no more without violating the positivity constraints of the pairwise transitions. The identifiability condition in Theorem \ref{identMTD} also offers an interpretation of the parameters in the MTD model. Specifically, the element ${\bf Z}^j_{mn}$ denotes the additive increase in probability that $x_i$ is in state $m$ given that $x_j$ is in state $n$. Furthermore, the $\gamma^j$ parameters now represent the total amount of probability mass in the full conditional distribution explained by categorical variable $x_j$, providing an interpretable notion of dependence in categorical time series. The mLTD model, however, does not readily suggest a simple and interpretable notion of dependence from the $\Zj$ matrix due to the non-linearity of the link function. The identifiability conditions are displayed pictorially in Figure \ref{ident_fig}.

Unfortunately, the necessary constraint set for identifiability specified in Theorem \ref{identMTD} is a non-convex set since the locations of the zero elements in each row of $\Zj$ are unknown. Naively, one could search over all possible locations for the  zero element in each row of each $\Zj$; however, this quickly becomes infeasible as both $m$ and $d$ grow.

Instead, we add a penalty term $\Omega({\bf Z})$, or prior, that biases the solution towards the uniqueness constraints. This regularization also aids convergence of optimization since the maximum likelihood solution without identifiability constraints is not unique. Letting $L_{\text{MTD}}({\bf Z}) = - \sum_{t = 1}^{T}\log \left(\zzero_{x_{it}} + \sum_{j = 1}^{p} \Zj_{x_{it} \,\, x_{j(t-1)}} \right)$ the regularized estimation problem is given by 
\begin{equation} \label{ident_mtd}
\begin{aligned}
&\underset{{\bf Z}, {\bf \gamma}}{\text{minimize}}  \,\,\, L_{\text{MTD}}({\bf Z}) + \lambda \Omega({\bf Z}) \,\,\,\, \\
&\text{subject to} \,\,\,\,\, {\bf 1}^T \Zj = \gamma_j{\bf 1}^T ,\,\,\, \Zj \geq 0 \,\, \forall j, \,\,\,\, {\bf 1}^T \gamma = 1 \,\,,\gamma \geq 0 .
\end{aligned}
\end{equation} 
\begin{ntheorem} \label{optid}
For any $\lambda > 0$ and $\Omega({\bf Z})$ that does not depend on $\zzero$ and is increasing with respect to the absolute value of entries in $\Zj$, the solution to the problem in Eq. (\ref{ident_mtd}) is contained in the set of identifiable MTD models described in Theorem \ref{mtd_ident}. 
\end{ntheorem}
Intuitively, by penalizing the entries of the $\Zj$ matrices, but not the intercept term, solutions will be biased to having the intercept contain the excess probability mass, rather than the $\Zj$ matrices. Thus, even with a very small penalty, we constrain the solution space to the set of identifiable models. Theorem \ref{optid} characterizes an entire \emph{class} of regularizers that enforce the identifiability constraints for MTD. As we explain in Section \ref{sec:MTDmodelsel}, a convenient choice for $\Omega({\bf Z})$ for our case coincides with a regularizer for selecting for Granger causality.  

\subsubsection{Identifiability for the mLTD model}
The non-identifiability of multinomial logistic models is also well-known, as is the non-identifiability of generalized linear models with categorical covariates. Combining the standard identifiability restrictions for both settings gives \cite{Agresti:2011}:
\begin{proposition}
(\cite{Agresti:2011}) Every mLTD has a unique parameterization such that first column and last row of $\Zj$ are zero for all $j$ and the last element of $\zzero$ is zero.
\label{identmLTD}
\end{proposition}
These conditions are displayed pictorially in Figure \ref{ident_fig}. Under the identifiability constraints for both MTD and mLTD models, at least one element in each row must be zero. For MTD this zero may be in any column, while for mLTD the zero may be placed in the first column of each row without loss of generality. For mLTD the last row of $\Zj$ must also be zero due to the logistic output (one category serves as the `baseline'); in MTD, instead, each column of $\Pj$ must sum to one.

\subsection{Granger Causality in MTD and mLTD}

Under the $\Zj$ MTD parameterization and the mLTD specification of Eq. (\ref{GLM}), we have the following simple result for Granger non-causality conditions:
\begin{proposition} \label{gc}
In both the MTD model of Eq. (\ref{MTDreparam}) and the mLTD model of Eq. (\ref{GLM}), time series $x_j$ is Granger non-causal for time series $x_i$ iff the columns of $\Zj$ are all equal.
\end{proposition}
Intuitively, if all columns of $\Zj$ are equal, the transition distribution for $x_{it}$ does not depend on $x_{j(t-1)}$. This result for mLTD and MTD models is analogous to the general Granger non-causality result for the slices of the conditional probability tensor being constant along the $x_{j(t-1)}$ mode being equal. Based on Proposition \ref{gc}, we might select for Granger non-causality by penalizing the columns of $\Zj$ to be the same. While this approach is potentially interesting, a more direct, stable method takes into account the conditions required for identifiability of the $\Zj$ under both models.

Under the identifiability constraints for both MTD and mLTD given in Theorems \ref{identMTD} and Proposition \ref{identmLTD}, respectively, then $x_j$ is Granger non-causal for $x_i$ iff $\Zj = 0$ (a special case of all columns being equal). For both MTD and mLTD models this fact follows from each row having at least one zero element; for all the columns to be equal as stated in Proposistion \ref{gc}, all elements in each row must also be equal to zero. Taken together, if we enforce the identifiability constraints, we may uniquely select for Granger non-causality by encouraging some $\Zj$ to be zero.

\section{Granger Causality Selection} \label{opt}
We now turn to procedures for inferring Granger non-causality statements from observed multivariate categorical time series.  In Section~\ref{ident}, we derived that if $\Zj=0$, then $x_j$ is Granger non-causal for $x_i$ in both MTD and mLTD models.  To perform model selection, we take a penalized likelihood approach and present a set of penalty terms that encourage $\Zj=0$ while maintaining convexity of the overall objective. 
The final parameter estimates automatically satisfy the identifiability constraints for MTD. We also develop analogous penalized criterion for selecting Granger causality in the mLTD model.

\subsection{Model selection in MTD}
\label{sec:MTDmodelsel}
We now explore penalties that encourage the $\Zj$ matrices to be zero.  Under the $\Pj, \gamma_j$ parameterization this is equivalent to encouraging the $\gamma_j$ to be zero. We first introduce an $L_0$ penalized problem in terms of the original $\gamma_j$ parameterization, and then show how convex relaxations of the $L_0$ norm on $\gamma_j$ lead to natural convex penalties on $\Zj$. Ideally, we would solve the penalized $L_0$ problem:
\begin{equation} \label{mtd_l0}
\begin{aligned}
& \underset{{\bf Z}, {\bf \gamma}}{\text{minimize}} \,\,\,  L_{\text{MTD}}({\bf Z}) + \lambda ||\gamma_{1:p}||_0 \,\,\,\, \\
& \text{subject to} \,\,\,\,\, {\bf 1}^T \Zj = \gamma_j{\bf 1}^T ,\,\,\, \Zj \geq 0 \,\, \forall j, \,\,\,\, {\bf 1}^T \gamma = 1 \,\,,\gamma \geq 0 
\end{aligned}
\end{equation} 
where $\lambda \geq 0$ is a regularization parameter and $||\gamma_{1:p}||_0$ is the $L_{0}$ norm over the $\gamma$ weights and the intercept weight $\gamma_0$ is not regularized. The $L_0$ penalty simply counts the number of non-zero $\gamma_j$, which is equivalent to the number of non-zero $\Zj$.  This results in a non-convex objective.  Instead, we develop alternative convex penalties suited to model selection in MTD.  Importantly, we require that any such penalty $\Omega({\bf Z})$ fall in the intersection of two penalty classes: 1) $\Omega({\bf Z})$ must be a convex relaxation to the $L_0$ norm in Problem (\ref{mtd_l0}) to promote sparse solutions and 2) $\Omega({\bf Z})$ must satisfy the conditions of Theorem \ref{optid} to ensure the final parameter estimates satisfy the MTD identifiability constraints. We propose and compare two penalties that satisfy these criteria.

Our first proposal is the standard $L_1$ relaxation, as in lasso regression, which simply sums the absolute values of $\gamma_j$.  This penalty encourages \emph{soft-thresholding}, where some estimated $\gamma_j$ are set exactly to zero while others are shrunk relative to the estimates from the unpenalized objective. Note that due to the greater than zero constraint, the $L_1$ norm on $\gamma_{1:d}$ is simply given by the sum $\sum_{j = 1}^{d} \gamma_j$. If $\gamma_0$ were included in the $L_{0}$ regularization, the $L_1$ relaxation would fail due to the $\gamma$ simplex constraints $1^T \gamma = 1 $, $\gamma \geq 0$ so the $L_1$ norm would always be equal to one over the feasible set \cite{pilanci:2012}. Our addition of an unpenalized intercept to the MTD model allows us to sidestep this issue and leverage the sparsity promoting properties of the $L_1$ penalty for model selection in MTD. The $L_1$ regularized MTD problem is thus given by 
\begin{equation} \label{mtd_l11}
\begin{aligned}
& \underset{{\bf Z}, {\bf \gamma}}{\text{minimize}} \,\,\,  L_{\text{MTD}}({\bf Z}) + \lambda \sum_{j = 1}^{d} \gamma_j \,\,\,\, \\
 &\text{subject to} \,\,\,\,\, {\bf 1}^T \Zj = \gamma_j{\bf 1}^T ,\,\,\, \Zj \geq 0 \,\, \forall j, \,\,\,\, {\bf 1}^T \gamma = 1 \,\,,\gamma \geq 0, 
\end{aligned}
\end{equation} 
Eq. (\ref{mtd_l11}) may be rewritten solely in terms of the $\Zj$ terms by noting that $\gamma_j = \frac{1}{m_j} 1^T \Zj 1$. Defining $\tilde{z}^T = (\text{vec}({\bf Z}_1)^T, \ldots, \text{vec}({\bf Z}_d)^T) $, and assuming $|\mathcal{X}_i| = m \,\, \forall i$ for simplicity of presentation, we can rewrite the MTD constraints as
\begin{align}
(I_d \otimes A) \tilde{z} = 0, \,\,\,\, {\bf 1}^T \tilde{z} = m, \,\,\, \tilde{z} \geq 0, \,\,\,\,\,  \,\,\,\,\, \nonumber
\end{align}
where
\begin{align}
A = \left(\begin{array}{c c c c c} 
{\bf 1}_m^T & - {\bf 1}_m^T & 0 & 0 & \ldots \\
0 & {\bf 1}_m^T & - {\bf 1}_m^T & 0 & \ldots \\
\ldots & \ldots & \ddots & \vdots & \vdots\\
0 & 0 & \ldots  & {\bf 1}_m^T & - {\bf 1}_m^T
\end{array} \right)
\label{Aeq}
 \end{align}
$I_d$ is a $d$-dimensional identity matrix. This gives the final penalized optimization problem only in terms of $\Zj$ as

\begin{equation} \label{mtd_l1}
\begin{aligned}
& \underset{{\bf Z}}{\text{minimize}} \,\,\,  L_{\text{MTD}}({\bf Z}) + \lambda \sum_{i = 1}^{d} \frac{1}{m} {\bf 1}^T \Zj {\bf 1} \,\,\,\, \\
 &\text{subject to} \,\,\,\,\,\, (I_d \otimes A) \tilde{z} = 0, \,\,\,\, {\bf 1}^T \tilde{z} = m, \,\,\, \tilde{z} \geq 0 
\end{aligned}
\end{equation} 
Writing the $L_1$ penalized problem in this form shows that the $L_1$ penalty increases with the absolute value of the entries in $\Zj$ and does not penalize the intercept, thus satisfying the conditions of Theorem \ref{optid}. As a result, the solution to the problem given in Eq. (\ref{mtd_l1}) automatically satisfies the MTD identifiability constraints. Furthermore, the solution will lead to Granger causality estimates since many of the $\Zj$ will be zero due to the $L_1$ penalty.

Another natural convex relaxation of the objective in Eq. (\ref{mtd_l0}) is given by a group lasso penalty on each $\Zj$. The relaxation is derived by writing the $L_0$ norm as a rank constraint in terms of $\Zj$, which then is relaxed to a group lasso. Specifically, assume all time series have the same number of categories, $m_j = m \,\,\, \forall j$. Due to the equality and greater than zero constraints
\begin{align}
||\gamma_{1:p}||_0 &= ||\left({\bf 1}^T \text{vec}({\bf Z}^1),\ldots, {\bf 1}^T \text{vec}({\bf Z}^p) \right)||_0 \nonumber\\
&= \text{rank}({\bf Q}^T {\bf Q}) \nonumber\\
&= \text{rank}({\bf Q}) \nonumber
\end{align}
where 
\begin{align}
{\bf Q} = \left(\begin{array}{c c c c}
\text{vec}({\bf Z}^1) & 0 & \ldots & 0 \\
0 &  \text{vec}({\bf Z}^2) & \ldots & 0 \\
0 & \ldots & \ddots & \vdots \\
0 & \ldots  & \ldots & \text{vec}({\bf Z}^p)
\end{array} \right). \nonumber
\end{align}
Thus we can use the nuclear norm on ${\bf Q}$ as a convex relaxation to $||\gamma_{1:p}||_0$. Furthermore, the nuclear norm of ${\bf Q}$ is given by the sum of $\Zj$ Froebenius norms,
\begin{align}
||{\bf Q} ||_{*} = \sum_{i = 1}^{p} ||\Zj||_F, \nonumber
\end{align} 
where $||.||_{*}$ is the nuclear norm and $||.||_{F}$ is the Froebenius norm. This group penalty gives the final problem
\begin{equation} \label{mtd_group}
\begin{aligned}
& \underset{{\bf Z}}{\text{minimize}} \,\,\,  L_{\text{MTD}}({\bf Z}) + \lambda \sum_{j = 1}^{d} ||\Zj||_{F} \,\,\,\, \\ 
&\text{subject to} \,\,\,\,\,\, (I_d \otimes A) \tilde{z} = 0, \,\,\,\, {\bf 1}^T \tilde{z} = m, \,\,\, \tilde{z} \geq 0. 
\end{aligned}
\end{equation} 
Here, we penalize $\Zj$ directly, rather than indirectly via $\gamma_j$. The group lasso penalty drives all elements of $\Zj$ to zero together, such that the optimal solution automatically selects some $\Zj$ to be all zero and others not. This effect naturally coincides with our conditions of Granger non-causality that \emph{all} elements of $\Zj = 0$. The group lasso penalty also satisfies the conditions of Theorem \ref{optid} since the $L_2$ norm is increasing with respect to each element in $\Zj$ and the intercept is not penalized. Thus, solutions to Problem (\ref{mtd_group}) automatically enforce the MTD identifiability constraints.

\subsection{Model selection in mLTD}
To select for Granger causality in the mLTD model, we add a group lasso penalty to each of the $\Zj$ matrices, analogously to Eq. \eqref{mtd_group}, leading to the following optimization problem:
\begin{equation} 
\begin{aligned}
& \underset{{\bf Z}}{\text{minimize}} \,\,\,\,\, \sum_{t = 1}^{T} \zzero_{x_{it}} + \sum_{j = 1}^{d} \Zj_{x_{it}, x_{j(t-1)}}  + \log \left(\sum_{x' \in \mathcal{X}_i} \exp\left(\zzero_{x'} + \sum_{j = 1}^{d} \Zj_{x', x_{j(t-1)}}\right)\right) + \lambda \sum_{j = 1}^{d} ||\Zj||_{F} \\ & \text{subject to} \,\,\,\,\, \Zj_{1:m_i,1} = 0, \Zj_{m_i ,1:m_j} = 0 \,\,\, \forall j.
\end{aligned}
\end{equation} 
For two categories, $m_i = 2 \,\,\, \forall i$, this problem reduces to sparse logistic regression for binary time series, which was recently studied theoretically \cite{Hall:2016}. As in the MTD case, the group lasso penalty shrinks some $\Zj$ entirely to zero thereby selecting for Granger non-causality. 

\section{Optimization}
For both penalized MTD and mLTD models we use proximal gradient based methods for optimization. For the mLTD model we perform gradient steps with respect to the mLTD likelihood followed by a proximal step with respect to the group lasso penalty. This leads to a gradient step of the smooth likelihood followed by separate soft group thresholding \cite{Parikh:2014} on each $\Zj$.  

For the MTD model, our proximal algorithm reduces to a projected gradient algorithm \cite{Parikh:2014}. Projected gradient algorithms take steps along the gradient of the objective, and then project the result onto the feasible region defined by the constraints. In comparison to other MTD optimization methods, our projected gradient algorithm under the $\Zj$ parameterization is guaranteed to reach the global optima of the MTD log-likelihood. The gradient of the regularized MTD model with respect to entries in $\Zj$ over the feasible set is given by
\begin{align} \label{mtd_grad}
\frac{d L}{d \Zj_{x^{'},x^{''}}} &= \sum_{t = 1}^{T} 1_{\{x_{it} = x^{'}, x_{j(t-1)} = x^{''}\}} \frac{1}{\zzero_{x_{it}} + \sum_{j = 1}^{p} \Zj_{x_{it},x_{j(t-1)}}}
+ \lambda \frac{d \Omega}{d \Zj_{x^{'},x^{''}}}.
\end{align}
For the $L_1$ norm, $\Omega({\bf Z})$ is not differentiable when an element in any $\Zj$ is zero. For the $L_2$ group norm, $\Omega({\bf Z})$ is not differentiable when \emph{every} element in at least one $\Zj$ is zero. However, the MTD constraints enforce that $\Zj \geq 0$. Since the point of non-differentiability for both $L_1$ and $L_2$ norms occurs when elements are identically zero, we modify the constraints so that $\Zj \geq \epsilon$ for some small $\epsilon$. This allows us to ignore non-differentiability issues, and instead take gradient steps directly along the penalized MTD objective. 

Following the notation from the end of Section \ref{sec:MTDmodelsel}, let the set $C = \{\tilde{z} | \tilde{z}\geq \epsilon, (I_d \otimes A) \tilde{z} = 0, 1^T\tilde{z} = m\}$ denote the modified MTD constraints with respect to the $\Zj$ parameterization. We perform projected gradient descent by taking a step along the regularized MTD gradient of Eq. (\ref{mtd_grad}) and then projecting the result onto $C$. Specifically, the algorithm iterates the following recursion starting at iteration $k = 0$
\begin{align}
\tilde{z}^{k + 1} = \mathcal{P}_{C}\left(\tilde{z}^{k} - \delta_k \frac{d L}{d \tilde{z}}\right),
\end{align}
where $\delta_k$ is a step size chosen by line search \cite{Parikh:2014}.  We have written the projected gradient steps in terms of the vectorized variables $\tilde{z}$, rather than the $\Zj$ matrices, for ease of presentation. The $\mathcal{P}_{C}(x)$ operation is the projection of a vector $x$ onto the modified MTD constraint set $C$:
\begin{equation*}
\begin{aligned}
& \underset{z}{\text{minimize}} \,\,\,\,\, ||z - x||_2^2 \\
& \text{subject to} \,\,\,\,\, z \geq \epsilon,\,\,\,\,\, (I_d \otimes A)z  = 0,\,\,\,\, {\bf 1}^T z = m.
\end{aligned}
\end{equation*}
This is a quadratic program and we use the the dual method \cite{goldfarb:1982} as implemented in the R quadratic programming package \emph{quadprog} \cite{turlach:2013}. However, we have found that this standard R solver scales poorly as the number of time series $d$ gets large. Instead, we have developed a fast projection algorithm based on Dykstra's splitting algorithm \cite{boyle:1986} that harnesses the particular structure of the constraint set for much faster projection, as described in Section \ref{dyk}. The full projected gradient algorithm for MTD is given in Algorithm \ref{projMTD}. 

\subsection{Dykstra's Splitting Algorithm for Projection onto the MTD Constraints} \label{dyk}
The set $C$ may be written as the intersection of two simpler sets: $C = S \cap B$, where $S$ is the simplex constraint set of the first column of each $\Zj$ matrix and the greater than zero constraint for all entries of $\Zj$. Specifically,
\begin{align}
S = \left\{ \{\Zj \in \mathbb{R}^{m \times m} \}_{j = 0}^d \bigg| \sum_{j = 0}^p \sum_{i = 1}^m \Zj_{1i} = 1, \Zj \geq 0  \forall j \right\}.
\end{align}
On the other hand, $B = \cup_{j = 1}^p B_j $, where $B_j$ is the constraint set that all columns in $\Zj$ sum to the same value:
\begin{align}
B_j = \left\{ \Zj \in \mathbb{R}^{m \times m} \bigg | A \,\, \text{vec}(\Zj) = {\bf 0} \right\},
\end{align}
where the matrix $A$ is given in Eq. (\ref{Aeq}).
Dykstra's algorithm alternates between projecting onto the simplex constraints $S$ and the equal column sums $B$ by iterating the following steps. Let $w^0 = x, u^0 = v^0 = 0$ and repeatedly update starting with iteration number $l = 0$:
\begin{itemize}
\item[] $y^l = \mathcal{P}_S(w^l + u^l)$
\item[] $u^{l + 1} = w^l + u^l - y^l$
\item[] $w^l = \mathcal{P}_B(y^l + v^l)$
\item[] $v^{l + 1} = y^l + v^l - w^l$
\end{itemize}
where $\mathcal{P}_S$ is the projection onto the set $S$ and $\mathcal{P}_B$ is the linear projection onto the set $B$. The $\mathcal{P}_S$ projection may be split into a simplex projection for the constraint $\sum_{j = 0}^d \sum_{i = 1}^m \Zj_{1i} = 1, \Zj_{1i} \geq 0 \,\,\, \forall i,j$ and a greater than zero constraint $\Zj_{ni} \geq 0 \,\,\, \forall i,j$ and $n > 1$. We perform the simplex projection in $(d m) \log (d m)$ time using the algorithm of \cite{duchi:2008} and the greater than zero projection is simply thresholding elements at zero and is performed in linear time. The $\mathcal{P}_B$ linear projection is performed separately for each $\Zj$:
\begin{align} \label{Bj}
\mathcal{P}_{B_j}(x) = \left(I - \left(A \left(A A^T\right)^{-1} A^T\right) \right) x
\end{align}
where $\left(I - \left(A \left(A A^T\right)^{-1} A^T\right) \right)$ may be precomputed so the per-iteration complexity for the full $B$ projection is $d m^4$ since $A$ is a $(m-1) \times m^2$ matrix.  Importantly, this projection scheme harnesses the structure of the constraint set by splitting the projections into components that admit fast and simple low-dimensional projections. The full projection algorithm is given in Algorithm \ref{zyk}.

\begin{figure} \label{projcomp}
\centering
\includegraphics[width=0.4\textwidth]{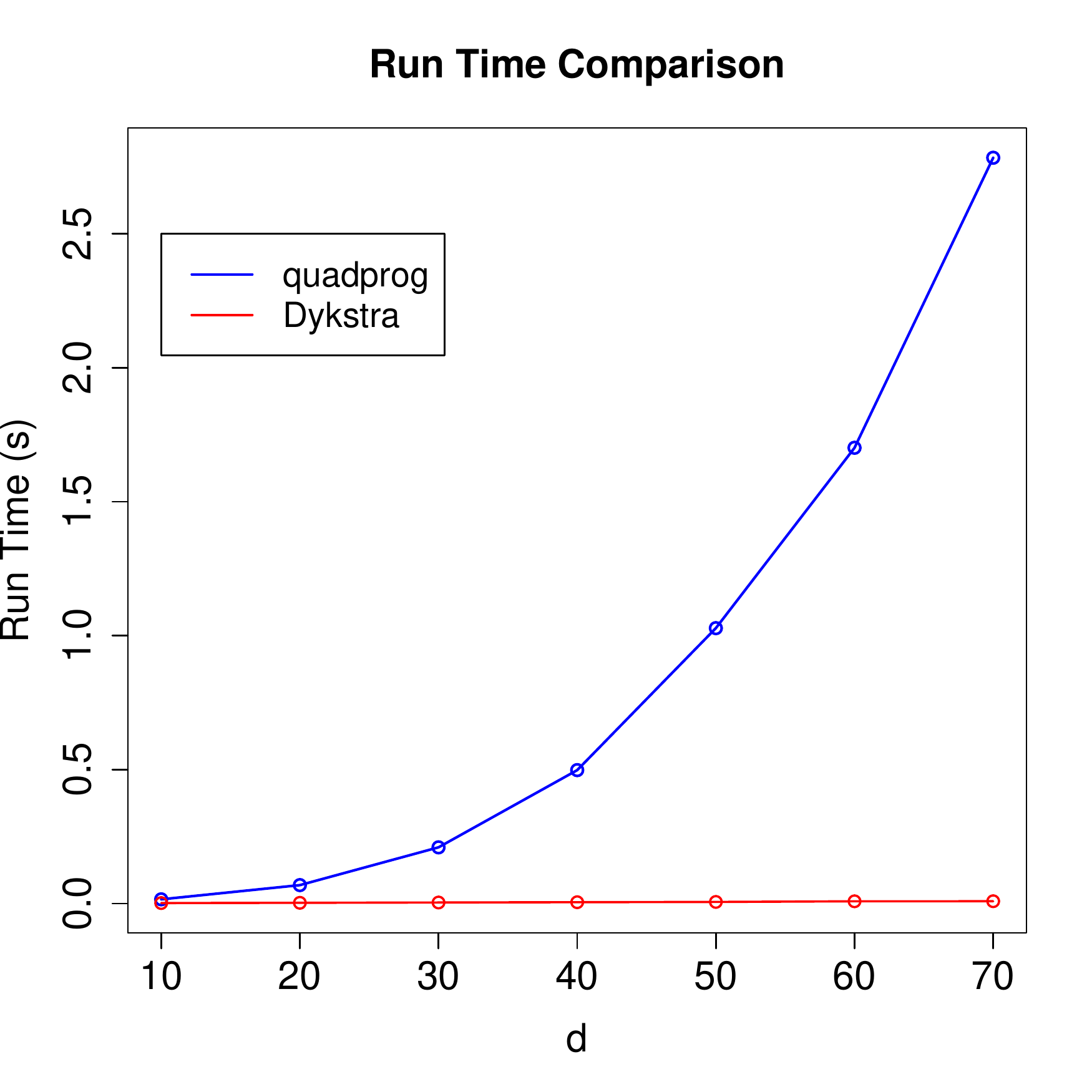}
\includegraphics[width=0.4\textwidth]{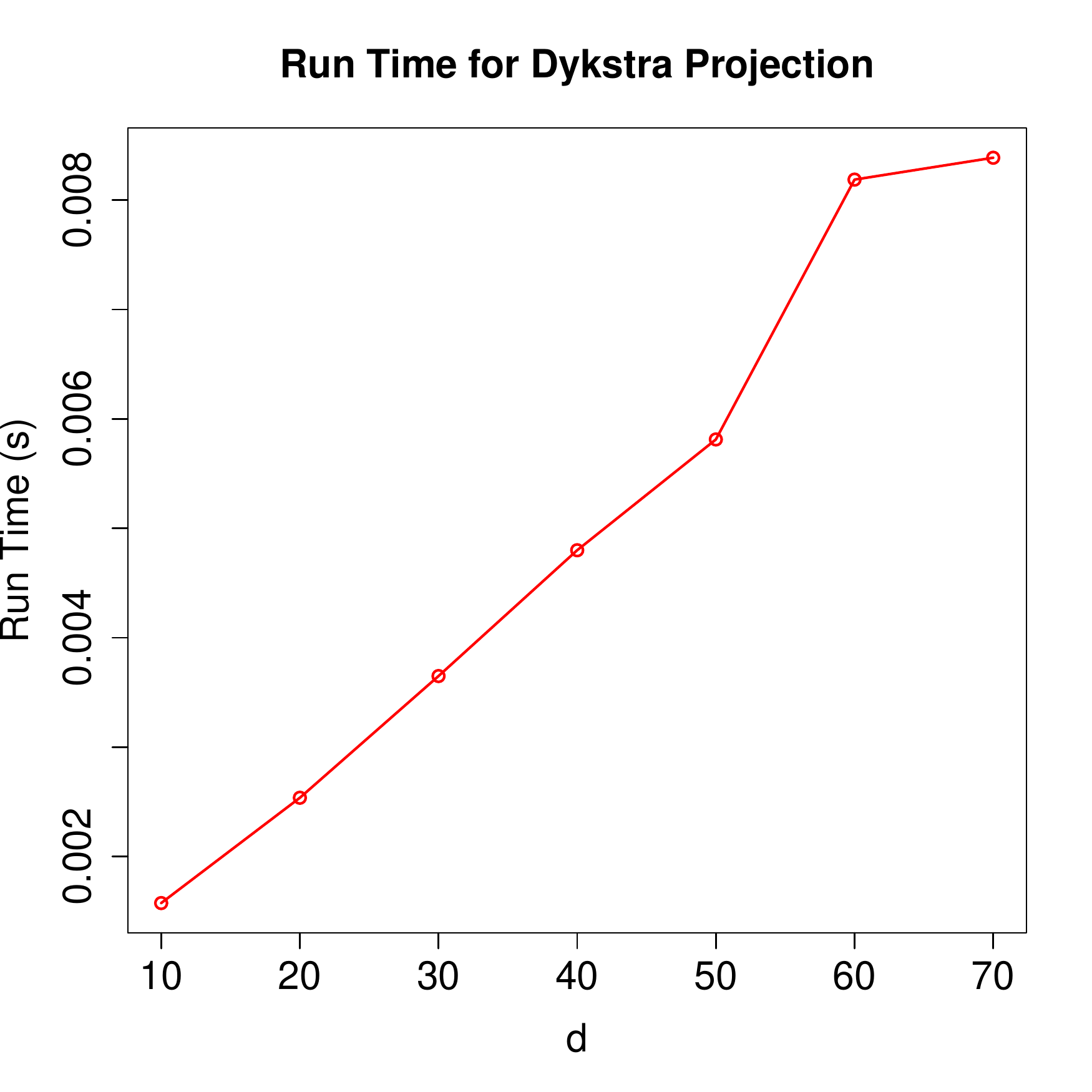}
\caption{(\emph{left}) A runtime comparison of the quadprog projection method and the Dykstra projection method on a range of time series dimensions. (\emph{right}) A zoom in of only the compute time of the Dykstra method.}
\end{figure}

We compare projection times of the Dykstra algorithm to the active set method of \cite{goldfarb:1982} implemented in the R package \emph{quadprog} \cite{turlach:2013}. The Dykstra projection for the MTD constraints was implemented in C++. Elements of $\Zj$ were drawn independently from a normal distribution with standard deviation $.7$ and then projected onto $C$. Average run times across 10 random realizations for $d \in (10,20,30,40,50,60)$ series and $m = 5$ categories are displayed in Figure \ref{projcomp}. The Dykstra algorithm was run until iterates changed by less than $10^{-10}$. For each run, the elementwise maximum difference between the Dykstra projection the \emph{quadprog} projection was always on the scale of $10^{-10}$.  Across this range of $d$ the \emph{quadprog} runtime appears to scale quadratically in $d$, with a total run time on the scale of seconds for $d \geq 20$. The Dykstra projection method, however, appears to scale near linearly in this range with run times on the order of milliseconds. We also performed experiments with differing standard deviations for the independent draws of $\Zj$ and the results were all very similar.

\subsection{Comparing model selection and optimization in MTD and mLTD}
Approaches to model selection in MTD and mLTD models are conceptually similar; both add regularizing penalties to enforce elements in $\Zj$ to zero. However, these two approaches differ in practice. We explore the differences in selecting for Granger causality between these two approaches via extensive simulations in Section \ref{experiments}.

Both MTD and mLTD models take gradient steps followed by a proximal operation. In the mLTD model this proximal operation is given by soft thresholding on the elements of $\Zj$. In the MTD optimization the proximal operation reduces to a projection onto the MTD constraint set. Importantly, due to the restricted domain of the MTD parameter set, the normally non-smooth penalty terms become smooth over the constraint set and we thus include them in the gradient step. In mLTD, the soft threshold proximal operation is performed in linear time while in MTD the projection is performed by iteratively using the Dykstra algorithm, where each step of the Dykstra algorithm is performed in log-linear time. 

\begin{algorithm}[H]
 \KwData{X}
 \KwResult{$\hat{\bf Z}$}
 Initialize ${\bf Z}^0 \,\,\ \forall j$ \;
 $k = 0$ \;
 \While{${\bf Z}^k$ not converged}{
  compute $\nabla L({\bf Z}^k)$ via Eq. (\ref{mtd_grad})\;
  determine $\gamma^k$ by line search \cite{Parikh:2014}\;
  ${\bf Z}^{k+1} = DykstraMTD\left({\bf Z}^k + \gamma^k \nabla L({\bf Z}^k)\right)$\;
  $k = k + 1$\;
  }
   \caption{Projected gradient algorithm for MTD using Dykstra projections.} \label{projMTD}
\end{algorithm}

\begin{algorithm}[H]
 \KwData{${\bf Z}$}
 \KwResult{$P_C({\bf Z})$}
 $z = \left(({\bf z}^{0})^T, vec({\bf Z}^1)^T, \ldots, vec({\bf Z}^p)^T\right)^T$ \;
 Let $S$ be the ordered indices of $z$ whose elements belong in the first column of some $\Zj, \,\,  j > 0$ or in ${\bf z}^0$ \;
 Let $(j)$ refer to ordered indices of $z$ whose elements belong to $\Zj$ $\forall j$. \;
 $w_0 = z$\;
 $u_0 = v_0 = 0$\;
 $l = 0$\;
 \While{$w^l$ not converged}{
  $y^l_{S} = SimplexProjection(w^l_S + p^l_S)$ via \cite{duchi:2008}\;
  $y^l_{\backslash S} = PositiveThreshold\left(w^l_{\backslash S} + u^l_{\backslash S}\right)$\;
  $u^{l+1} = w_l + u_l - y_l$\;
  $w^k_{(0)} = y^l_{(0)} + v^l_{(0)}$\;
  \For{j = 1:p} {
    $w^l_{(j)} = P_{B_j}\left(y^l_{(j)} + v^l_{(j)}\right)$ via Eq. (\ref{Bj})\;
  }
  $v^{(l + 1)} = y^l + q^l - w^l$\;
  $l = l + 1$\;
  }
   \caption{\emph{DykstraMTD}: Zykstra algorithm for projection onto the MTD constraints.} \label{zyk}
\end{algorithm}

\section{Experiments} \label{experiments}
\subsection{Simulation Set Up}
We perform a set of simulation experiments to compare the MTD and mLTD 
model selection methods. Specifically, we compare the MTD group lasso, $L_1$-MTD, and mLTD group lasso methods on simulated categorical time series generated first from a sparse MTD model. We find that the group lasso MTD outperforms the MTD $L_1$ and thus only compare MTD group lasso and mLTD group lasso on two further simulated scenarios: a sparse mLTD model and a sparse latent vector autoregressive model (VAR) with quantized outputs. For all experiments we consider time series of length $T \in (200,400)$, dimension $d \in (15,25)$, and number of categories $m \in (2,3,4,5,6)$. We first explain the details of each simulation condition and then discuss the results.
\paragraph{Sparse MTD}
For the MTD model, we randomly generate parameters by $\gamma_{ij} \sim \frac{z_{ij} \phi_{ij} }{\sum_{l = 1}^p z_{il} \phi_{il}}$ where $\phi_i \sim \text{Dirichlet}(\alpha)$ and $z_{ij} \sim \text{Binomial}(\delta)$. We let $\delta = .15, \alpha = 5$. Columns of ${\bf Z}^{ij}$ are generated according to ${\bf Z}^{ij}_{:l} \sim \text{Dirichlet}(\gamma)$ with $\gamma = .7$. (Note that here we have added a superscript $i$ to ${\bf Z}$ to specifically indicate the $j$ to $i$ interaction, whereas previously we dropped the $i$ index for notational simplicity by assuming we were just looking at the series $i$ term.) To ensure that the columns are not close to identical in ${\bf Z}^{ij}$ (which would imply Granger non-causality), ${\bf Z}^{ij}$ is sampled until the average total variation norm between the columns is greater than some tolerance $\rho$.  This ensures that non-causality occurs only due to which $\Zj$ are zero, and not due to equal columns in the simulation. For our simulations, we set $\rho = .3$. A lower value of $\rho$ makes it more difficult to learn the Granger causality graph since some true interactions might be extremely weak. 
\paragraph{Sparse mLTD} For the mTLD model, the nonzero ${\bf Z}^{ij}$ parameters are generated by ${\bf Z}^{ij}_{lk} \sim z_{ij} N(0,\sigma_Z^2)$ where $z_{ij} \sim \text{Binomial}(\delta)$ with $\delta = .15$. 
\paragraph{Sparse Latent VAR} To examine data generated from neither of the models considered, we simulate data from a continuous time series $y_t \in \mathbb{R}^p$ according to a sparse VAR(1):
\begin{align}
y_t = A y_{t-1} + \epsilon_t \nonumber
\end{align}
where $\epsilon_t \sim N(0,\sigma^2 I_p)$. The sparse matrix $A$ is generated by first sampling entries $B_{ij} \sim N(0,\sigma_A^2)$ and then setting $A_{ij} = B_{ij}z_{ij}$, where $z_{ij} \sim \text{Binomial}(\delta)$ with $\delta = .15$. We then quantize each dimension, $y_{ti}$, into $m$ categories to create a categorical time series $x_{ti}$. For example, when $m = 3$, $x_{ti} = 1$ if $y_{ti}$ is in the $(0,.33)$ quantile of $\{y_{1i}, \ldots y_{Ti} \}$, and so forth. 

\subsection{Simulation Results}
For all methods - MTD $L_1$, MTD group lasso, and mLTD group lasso - we compute the area under the ROC curve between the true Granger causality graph and the sparse graph that results when varying $\lambda$ across a range of values. 

\begin{figure}
\centering
\includegraphics[width=.7\textwidth]{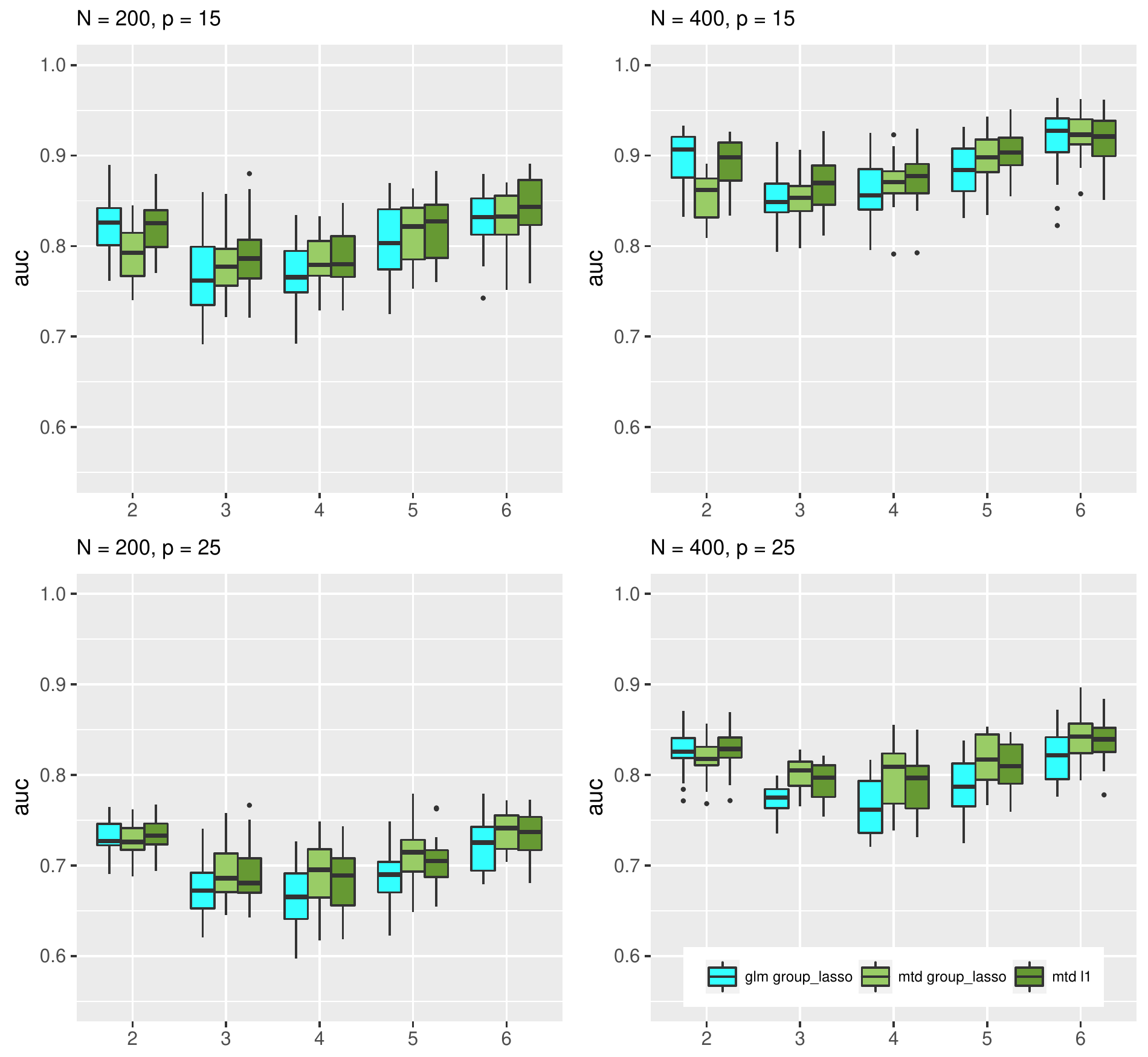}
\caption{AUC for data generated by a sparse MTD process. Boxplots over 20 simulation runs.}
 \label{AUC_MTD}
\end{figure}
\begin{figure}
\centering
\includegraphics[width=.7\textwidth]{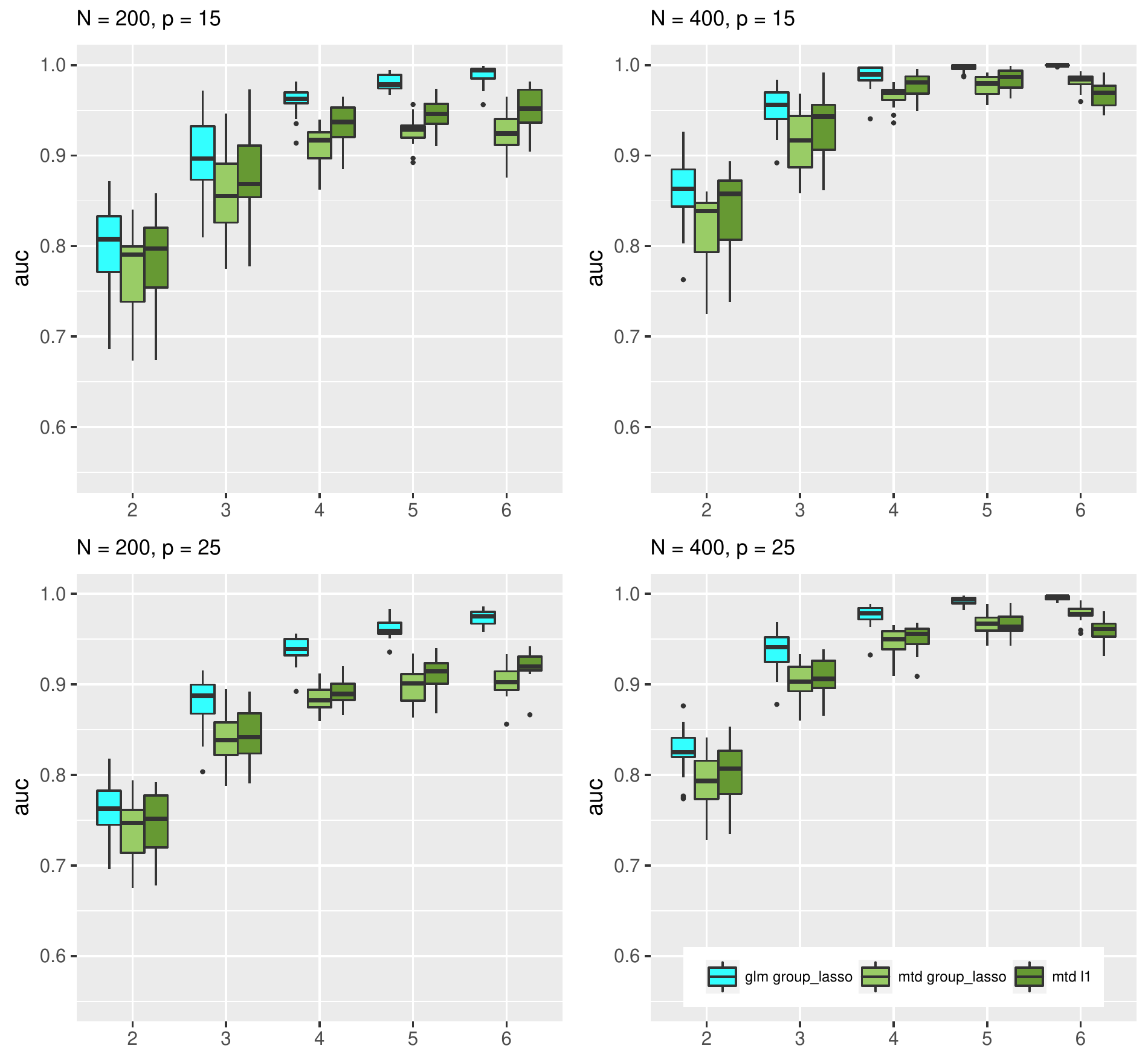}
\caption{AUC for data generated by a sparse latent mLTD process. Boxplots over 20 simulation runs.}
 \label{AUC_mLTD}

\end{figure}
\begin{figure}
\centering
\includegraphics[width=.7\textwidth]{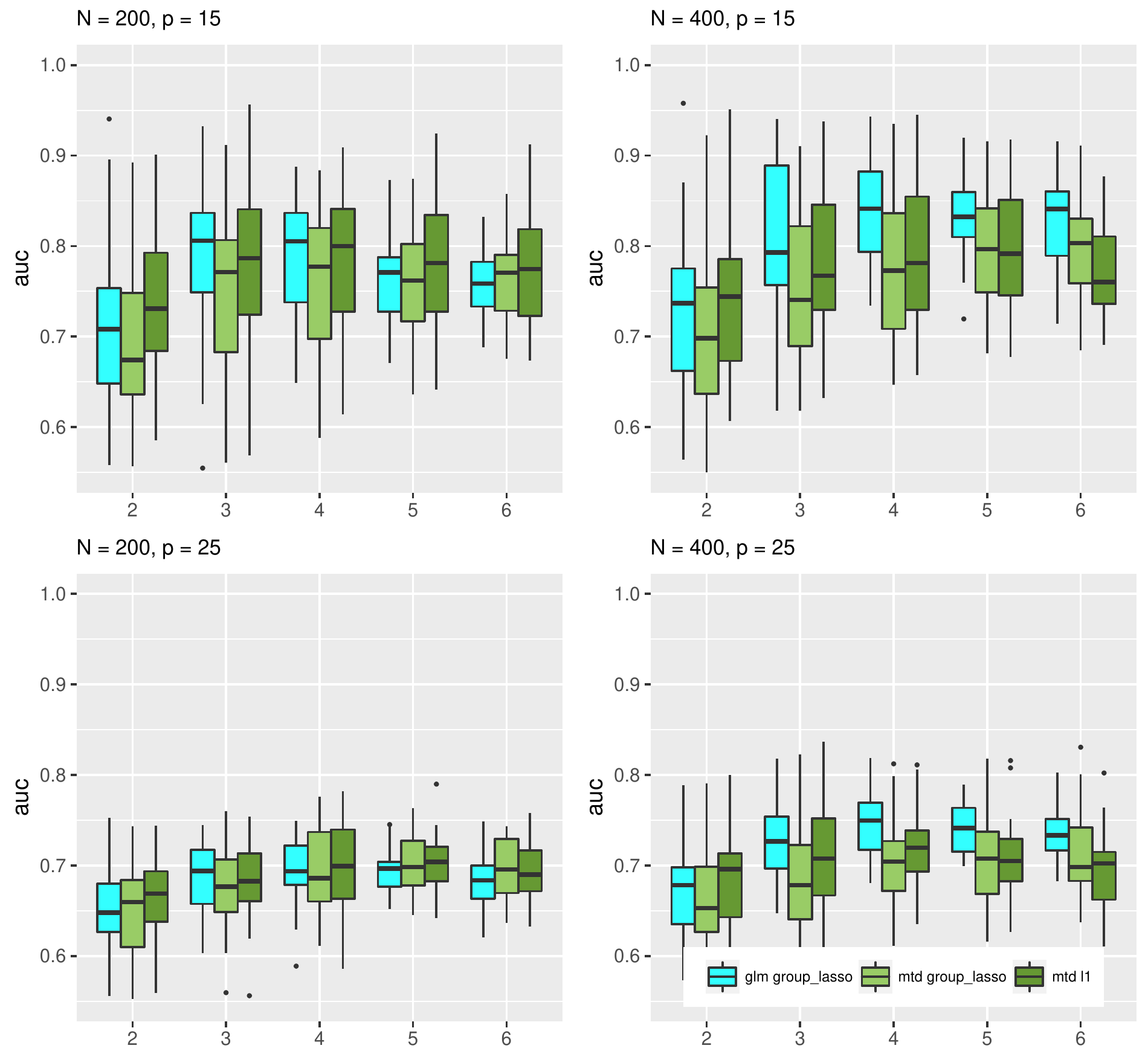}
\caption{AUC for data generated by a sparse latent VAR process. Boxplots over 20 simulation runs.}
 \label{AUC_VAR}
\end{figure}

The results are displayed as histograms across all simulation runs in Figures \ref{AUC_MTD}, \ref{AUC_mLTD}, and \ref{AUC_VAR} for the categorical time series generated by MTD, mLTD, and latent VAR, respectively. We note that the mLTD group lasso model performs best when the data are generated from a mLTD, and likewise the MTD group lasso performs best when the data are generated from a MTD. Furthermore, the MTD $L_1$ estimator tends to outperform the MTD group lasso across most settings. Interestingly, for data generated from mLTD we see improved performance as a function of the number of categories $m$ for all $n$ and $d$ settings, while for MTD performance starts high, dips and goes back up with increasing $m$. This is probably due to the simulation conditions, as in both MTD and mLTD models Granger causality can be quantified as the difference between the columns of $\Zj$. When there are more categories, there is higher probability under our simulation conditions that there will be some columns with large deviation from other columns in $\Zj$. This leads to improved Granger causality detection when it exists.

In the latent VAR simulation, MTD group and mLTD group perform similarly in the $T = 200$ simulation condition, but mLTD consistently outperforms MTD in the $T = 400$ case. Taken together, though, both methods perform comparably. There is also evidence of improved performance for both MTD and mLTD methods as the quantization of the latent VAR processes becomes finer. For the MTD model the average AUC increases rougly monotonically with quantization level, though for the mLTD average performance appears to peak at $m = 4$ categories and then levels off or slightly declines. When the quantization is too coarse, say for $m = 2$ or $m = 3$, some Granger causality interactions may become hard to detect since there is much less information about the underlying VAR processes contained in the quantized series. 

As expected, across all simulation conditions and estimation methods increasing the sample size $T$ leads to improved performance while increasing the dimension $d$ worsens performance.

\section{Music Data Analysis}
We analyze Granger causality connections in the `Bach Choral Harmony' data set available at the UCI machine learning repository \cite{Lichman:2013} (\url{https://archive.ics.uci.edu/ml/datasets/Bach+Chorales}). This data set has been used previously in \cite{radicioni:2010,esposito:2009}. The data set consists of 60 chorales for a total of 5665 time steps. At each time step 15 unique discrete events are recorded. There are 12 harmony notes, $\{\text{C, C\#, D, , D\#, E, F, G, G\#, A , A\#, B}\}$, that take values either `on' (played) or `off' (not played), i.e. $x_{tj} \in \{0,1\}$ for $j \in \{1, \ldots, 12\}$. There is one `meter' category taking values in $\{1, \ldots, 5\}$, where lower numbers indicate less accented events and higher numbers higher accented events. There is also the `pitch class of the base note', taking 12 different values and a `chord' category. We group all chords that occur less than 200 times into one group, giving a total of 12 chord categories. 

\begin{figure}
\label{bach_graph}
\centering
\includegraphics[width = .5\textwidth]{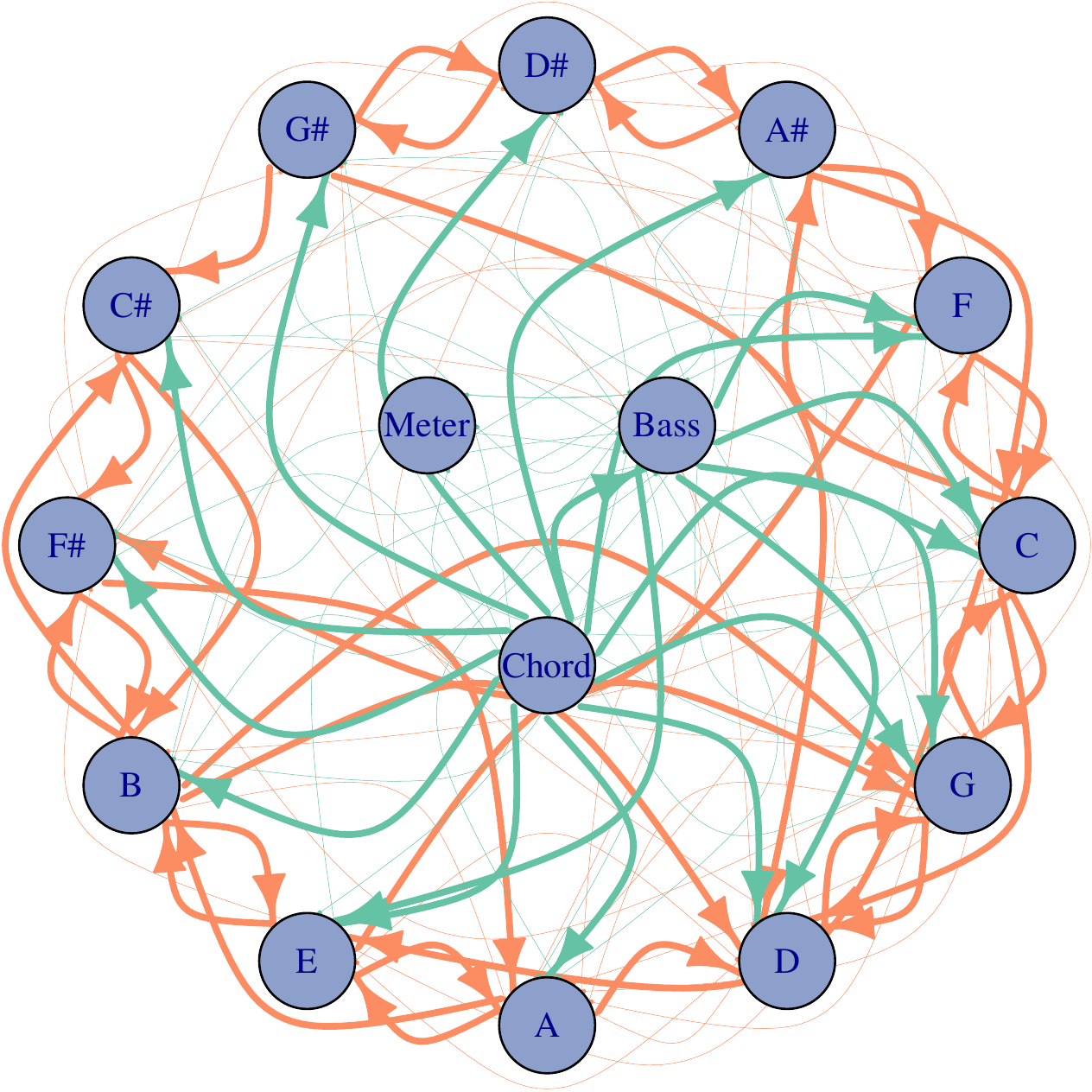}
\caption{The Granger causality graph for the `Bach Choral Harmony' data set using the penalized MTD method. The harmony notes are displayed around the edge in a circle corresponding to the circle of fifths. Orange links display directed interactions between the harmony notes while green links display interactions to and from the `bass', `chord', and `meter' variables.}
\end{figure}

We apply the sparse MTD model for Granger causality selection and choose the tuning parameter $\lambda$ by a five-fold cross validation over a grid of $\lambda$ values. We threshold the $\gamma$ weights at .01 and plot the estimated resulting Granger causality graph in Figure \ref{bach_graph}. For further interpretability we bold all edges with $\gamma$ weight magnitudes greater than .06. As mentioned in Section \ref{mtd_ident_s}, the MTD model is much more appropriate than the mLTD model for this type of exploratory Granger causality analysis: The $\gamma$ weights intuitively describe the amount of probability mass that is accounted for in the conditional probability table, giving an intuitive notion of dependence between categorical variables. In the mLTD model, however, it is not clear how to define strength of interaction and dependence given a set of estimated $\Zj$ parameters due to the non-linearity of the softmax function.

The harmony notes in the graph are displayed in a circle corresponding to the circle of fifths. The circle of fifths is a sequence of pitches where the next pitch in the circle is found seven semitones higher or lower, and it is a common way of displaying and understanding relationships between pitches in western classical music. Plotting the graph in this way shows substantially higher connections with respect to sequences on this circle. For example, moving both clockwise and counter-clockwise around the circle of fifths we see strong connections between adjacent pitches, and in some cases strong connections between pitches that are two hops away on the circle of fifths. Strong connections to pitches far away on the circle of fifths are much rarer. Together, this indicates that in these chorales there is strong dependence in time between pitches moving in both the clockwise and counter-clockwise direction on the circle of fifths. 

We also note that the `chord' category has very strong outgoing connections implying it has strong Granger causality selection with all harmony pitches. This result is intuitive, as it implies that there is strong dependence between what chord is played at time step $t$ and what harmony notes are played at time step $t + 1$. The bass pitch is also influenced by `chord' and tends to both influence and be influenced by most harmony pitches. Finally, we note that the `meter' category has much fewer and weaker incoming and outgoing connections, capturing the intuitive notion that the level of accentuation of certain notes does not really relate to what notes are played.

We also performed a connectivity analysis using the penalized mLTD model. However, the mLTD model presents some extra difficulties. Importantly, due to the non-linearity of the softmax function there is not as an intuitive interpretation of `link strength' between two categorical variables in mLTD as there is in the MTD model. For this reason, it is not clear how to define the strength of interaction and dependence given a set of estimated $\Zj$ parameters. We chose to use the normalized $L_2$ norm of each $\Zj$ matrix, $\frac{||Z^i_j||}{\sqrt{m_i} \sqrt{m_j}}$, as a measure of connection strength in the mLTD model. However, this metric does not have a direct interpretation with respect to the conditional probability tensor. Due to these interpetational difficulties we present the results of the mLTD Bach analysis in the Appendix. We note here that the final graph shows some of the structure of the MTD analysis, strong connections between chord and the harmony notes and some strong connections between notes on the circle of fifths. However, in general, the resulting graph is much less sparse and interpretable than the MTD graph.




\section{Discussion}
We have proposed a novel convex framework for the MTD model as well as two penalized estimation strategies that simultaneously promotes sparsity in Granger causality estimation and constrain the solution to an identifiable space.
We have also introduced the mLTD model as a baseline for multivariate categorical time series that although straightforward, has not been explored in the literature. Novel identifiability conditions for the MTD have been derived and compared to those for the mLTD model. For optimization, we have developed a novel projected gradient algorithm for the MTD model that harnesses the new convex formulation. We also develop a novel Dykstra projection method to quickly project onto the MTD constraint set, allowing the MTD model to scale to much higher dimensions. Our experiments demonstrate the utility of both the MTD and mLTD model for inference of Granger causality networks from categorical time series, even under model misspecification. 


There are a number of potential directions for future work. Since we have formulated both MTD and mLTD models as convex problems, the general theory for high dimensional estimators based on convex losses \cite{Negahban:2009} may be leveraged to prove consistency of both models. Recently, \cite{hallE:2016} established consistency of high dimensional autoregressive GLMs with univariate natural parameters for each series. An interesting direction would be to combine these general techniques for dealing with dependent observations with those of \cite{Negahban:2009} to derive rates for both the MTD and mLTD models. 

Further theoretical comparison between mLTD and MTD is also important. For example, to what extent may a mLTD distribution be represented by an MTD one, and vice versa; or, to what extent are both models consistent for Granger causality estimation under model misspecification. Our simulations results suggest that both methods perform well under model misspecification but more general theoretical results are certainly needed.

It would also be interesting to explore other regularized MTD objectives, such as the nuclear norm on $\Zj$ when the number of categories per time series is large. This penalty would both select for sparse dependencies while simultaneously share information about transitions within each $\Zj$. Another possibility includes the hierarchical group lasso over lags for higher order Markov chains, as in \cite{Nicholson:2014} for VARs, to automatically obtain the order of the Markov chain. Overall, the methods presented herein open up many new opportunities for analyzing multivariate categorical time series both in practice and theoretically.
\paragraph{\bf Acknowledgments}  This work was supported in part by
ONR Grant N00014-15-1-2380 and NSF CAREER Award IIS-1350133. AT was
partially funded by an IGERT fellowship. AS acknowledges the support from NSF grants  DMS-1161565 \& DMS-1561814 and NIH grants 1K01HL124050-01 \& 1R01GM114029-01.
\bibliography{cat}

\begin{thebibliography}{10}

\bibitem{Granger:1980}
Clive~WJ Granger.
\newblock Testing for causality: a personal viewpoint.
\newblock {\em Journal of Economic Dynamics and control}, 2:329--352, 1980.

\bibitem{Han:2013}
Fang Han, Huanran Lu, and Han Liu.
\newblock A direct estimation of high dimensional stationary vector
  autoregressions.
\newblock {\em arXiv preprint arXiv:1307.0293}, 2013.

\bibitem{Shojaie:2010}
Ali Shojaie and George Michailidis.
\newblock Discovering graphical {G}ranger causality using the truncating lasso
  penalty.
\newblock {\em Bioinformatics}, 26(18):i517--i523, 2010.

\bibitem{Zhou:2013}
Ke~Zhou, Hongyuan Zha, and Le~Song.
\newblock Learning social infectivity in sparse low-rank networks using
  multi-dimensional {H}awkes processes.
\newblock In {\em Proceedings of the Sixteenth International Conference on
  Artificial Intelligence and Statistics}, pages 641--649, 2013.

\bibitem{Hall:2016}
E.~C. {Hall}, G.~{Raskutti}, and R.~{Willett}.
\newblock Inference of high-dimensional autoregressive generalized linear
  models.
\newblock {\em ArXiv e-prints}, May 2016.

\bibitem{qiu:2015}
Huitong Qiu, Sheng Xu, Fang Han, Han Liu, and Brian Caffo.
\newblock Robust estimation of transition matrices in high dimensional
  heavy-tailed vector autoregressive processes.
\newblock In {\em Proceedings of the 32nd International Conference on Machine
  Learning (ICML-15)}, pages 1843--1851, 2015.

\bibitem{doshi:2011}
Finale Doshi, David Wingate, Josh Tenenbaum, and Nicholas Roy.
\newblock Infinite dynamic bayesian networks.
\newblock In {\em Proceedings of the 28th International Conference on Machine
  Learning (ICML-11)}, pages 913--920, 2011.

\bibitem{Raftery:1985}
Adrian~E Raftery.
\newblock A model for high-order {M}arkov chains.
\newblock {\em Journal of the Royal Statistical Society. Series B
  (Methodological)}, pages 528--539, 1985.

\bibitem{Nicolau:2014}
Jo{\~a}o Nicolau.
\newblock A new model for multivariate {M}arkov chains.
\newblock {\em Scandinavian Journal of Statistics}, 41(4):1124--1135, 2014.

\bibitem{Ching:2002}
Wai-Ki Ching, Eric~S Fung, and Michael~K Ng.
\newblock A multivariate {M}arkov chain model for categorical data sequences
  and its applications in demand predictions.
\newblock {\em IMA Journal of Management Mathematics}, 13(3):187--199, 2002.

\bibitem{berchtold:2002}
Andr{\'e} Berchtold and Adrian~E Raftery.
\newblock The mixture transition distribution model for high-order {M}arkov
  chains and non-{G}aussian time series.
\newblock {\em Statistical Science}, pages 328--356, 2002.

\bibitem{Zhu:2010}
Dong-Mei Zhu and Wai-Ki Ching.
\newblock A new estimation method for multivariate {M}arkov chain model with
  application in demand predictions.
\newblock In {\em Business Intelligence and Financial Engineering (BIFE), 2010
  Third International Conference on}, pages 126--130. IEEE, 2010.

\bibitem{Berchtold:2001}
Andre Berchtold.
\newblock Estimation in the mixture transition distribution model.
\newblock {\em Journal of Time Series Analysis}, 22(4):379--397, 2001.

\bibitem{bahadori:2013}
Mohammad~Taha Bahadori, Yan Liu, and Eric~P Xing.
\newblock Fast structure learning in generalized stochastic processes with
  latent factors.
\newblock In {\em Proceedings of the 19th ACM SIGKDD international conference
  on Knowledge discovery and data mining}, pages 284--292. ACM, 2013.

\bibitem{kedem:2005}
Benjamin Kedem and Konstantinos Fokianos.
\newblock Regression models for categorical time series.
\newblock {\em Regression Models for Time Series Analysis}, pages 89--137,
  2005.

\bibitem{Lebre:2008}
Sophie L{\`e}bre and Pierre-Yves Bourguignon.
\newblock An {EM} algorithm for estimation in the mixture transition
  distribution model.
\newblock {\em Journal of Statistical Computation and Simulation},
  78(8):713--729, 2008.

\bibitem{boyd:2004}
Stephen Boyd and Lieven Vandenberghe.
\newblock {\em Convex optimization}.
\newblock Cambridge university press, 2004.

\bibitem{Agresti:2011}
Alan Agresti and Maria Kateri.
\newblock {\em Categorical data analysis}.
\newblock Springer, 2011.

\bibitem{pilanci:2012}
Mert Pilanci, Laurent~E Ghaoui, and Venkat Chandrasekaran.
\newblock Recovery of sparse probability measures via convex programming.
\newblock In {\em Advances in Neural Information Processing Systems}, pages
  2420--2428, 2012.

\bibitem{Parikh:2014}
Neal Parikh and Stephen~P Boyd.
\newblock Proximal algorithms.
\newblock {\em Foundations and Trends in optimization}, 1(3):127--239, 2014.

\bibitem{goldfarb:1982}
Donald Goldfarb and Ashok Idnani.
\newblock Dual and primal-dual methods for solving strictly convex quadratic
  programs.
\newblock In {\em Numerical Analysis}, pages 226--239. Springer, 1982.

\bibitem{turlach:2013}
BA~Turlach and A~Weingessel.
\newblock quadprog {R} package. available online, 2013.

\bibitem{boyle:1986}
James~P Boyle and Richard~L Dykstra.
\newblock A method for finding projections onto the intersection of convex sets
  in hilbert spaces.
\newblock In {\em Advances in order restricted statistical inference}, pages
  28--47. Springer, 1986.

\bibitem{duchi:2008}
John Duchi, Shai Shalev-Shwartz, Yoram Singer, and Tushar Chandra.
\newblock Efficient projections onto the l 1-ball for learning in high
  dimensions.
\newblock In {\em Proceedings of the 25th international conference on Machine
  learning}, pages 272--279. ACM, 2008.

\bibitem{Lichman:2013}
M.~Lichman.
\newblock {UCI} machine learning repository, 2013.

\bibitem{radicioni:2010}
Daniele~P Radicioni and Roberto Esposito.
\newblock Breve: An hmperceptron-based chord recognition system.
\newblock In {\em Advances in Music Information Retrieval}, pages 143--164.
  Springer, 2010.

\bibitem{esposito:2009}
Roberto Esposito and Daniele~P Radicioni.
\newblock Carpediem: Optimizing the viterbi algorithm and applications to
  supervised sequential learning.
\newblock {\em Journal of Machine Learning Research}, 10(Aug):1851--1880, 2009.

\bibitem{Negahban:2009}
Sahand Negahban, Bin Yu, Martin~J Wainwright, and Pradeep~K Ravikumar.
\newblock A unified framework for high-dimensional analysis of {M}-estimators
  with decomposable regularizers.
\newblock In {\em Advances in Neural Information Processing Systems}, pages
  1348--1356, 2009.

\bibitem{hallE:2016}
Eric~C Hall, Garvesh Raskutti, and Rebecca Willett.
\newblock Inference of high-dimensional autoregressive generalized linear
  models.
\newblock {\em arXiv preprint arXiv:1605.02693}, 2016.

\bibitem{Nicholson:2014}
W.~B. {Nicholson}, J.~{Bien}, and D.~S. {Matteson}.
\newblock Hierarchical vector autoregression.
\newblock {\em ArXiv e-prints}, December 2014.

\end{thebibliography}
\bibliographystyle{unsrt}
\section{Appendix}

\subsection{mLTD Bach Analysis}
\begin{figure}
\label{bach_graph_app}
\centering
\includegraphics[width = .5\textwidth]{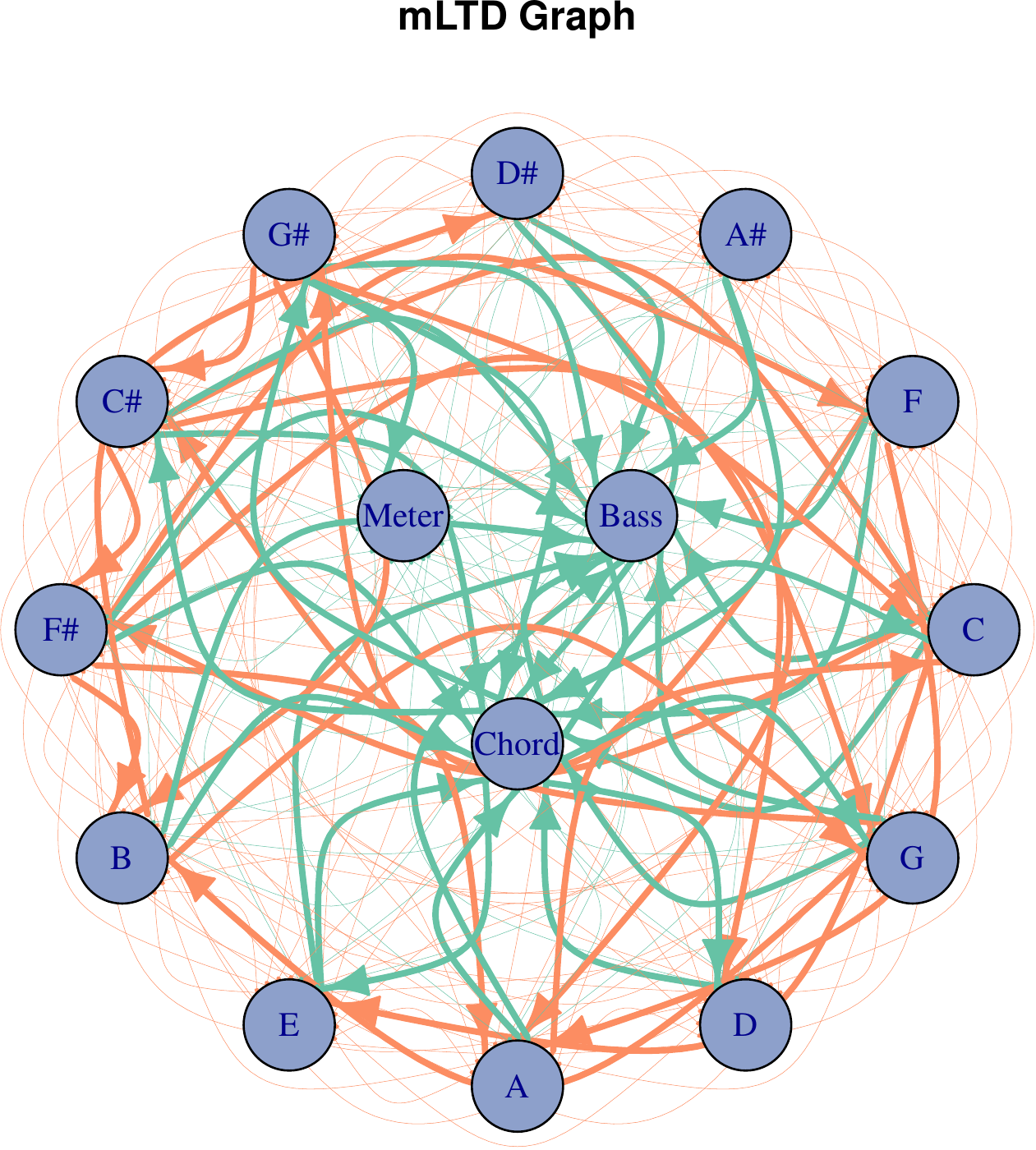}
\caption{The Granger causality graph for the `Bach Choral Harmony' data set using the mLTD method. The harmony notes are displayed around the edge in a circle corresponding to the circle of fifths. Orange links display directed interactions between the harmony notes while green links display interactions to and from the `bass', `chord', and `meter' variables.}
\end{figure}

For the mLTD Bach analysis we performed a 5-fold cross validation to select the $\lambda$ tuning parameter then thresholded the final connection weights, given by the standardised $L_2$ norm of $\Zj$, at .01, as in the MTD case. First, we note that the final mLTD model is much less sparse than the MTD case with only 5 total zero weights. We display the final graph in Figure \ref{bach_graph_app}, where, for interpretability, we bold edges with total weight greater than .45. In this graph there are strong connections in the counter clockwise direction between G\#, C\#, F\#, and B. However, the other connections on the circle of fifths are relatively weaker, and there are many more connections between notes far away on the circle of fifths. The mLTD graph also shows that the chord note both affects and is affected by many harmony notes. Furthermore, we see that the bass category is effected by most harmony notes as well. Overall, however, this graph is much less interpretable than the mTD graph and fails to find the full circle of fifths structure.

\subsection{Proofs}

\paragraph{Proof of Proposition \ref{gc}}
If the columns of $\Zj$ are all equal then for all fixed values of $x_{\backslash j(t-1)}$ the conditional distribution is the same for all values of $x_{j(t-1)}$. If one column is different then the conditional distribution for all values of $x_{\backslash j(t-1)}$ will depend on $x_{j(t-1)}$.
\paragraph{Proof of Theorem \ref{mtd_ident}}
Let ${\bf Z}$ be the parameter set for an MTD model. For each $\Zj$ let the vector $\alpha_j$ be the minimal element in each row. Let $\tilde{\Zj} = \Zj - \alpha_j$ and $\tilde{z} = z + \sum_{j = 1}^p \alpha_j$. This $\tilde{\bf Z}$ gives the same MTD
distribution as ${\bf Z}$.  

Suppose two parameter sets ${\bf X}$ and ${\bf Y}$ provide the same MTD distribution. Let $\tilde{\bf X}$ be the unique reduction of ${\bf X}$ and $\tilde{\bf Y}$ of ${\bf Y}$. Suppose $\tilde{\bf Y} \neq \tilde{\bf X}$. There must exist some $j$ and some row $k$ such that $\tilde{\bf X}^j_{k:} \neq \tilde{\bf Y}^j_{k:}$. Let $l_X$ be the index such that $\tilde{\bf X}^j_{kl} = 0$ and likewise for $l_Y$. 

If $l_X = l_Y$, let $l'$ be an index such that $\tilde{\bf X}^j_{kl'} \neq \tilde{\bf Y}^j_{kl'}$. Let $x_{\setminus j(t-1)}$ be fixed arbitrarily. The value of
\begin{align}
p_X(x_t = k|x_{\setminus j (t-1)}, x_{j(t-1)} = l') \nonumber \\
- p_X(x_t = k|x_{\setminus j (t-1)}, x_{j(t-1)} = l_X) &= \tilde{\bf X}^j_{kl'} \nonumber \\
&\neq \tilde{\bf Y}^j_{kl'} \nonumber \\
p_Y(x_t = k|x_{\setminus j (t-1)}, x_{(t-1)j} = l') \nonumber \\ - p_Y(x_t = k|x_{\setminus j(t-1)}, x_{(t-1)j} = l_Y) &= \nonumber
\end{align} 
showing the MTD distributions parametrized by ${\bf X}$ and ${\bf Y}$ are not the same.

If $l_X \neq l_Y$, then
\begin{align}
p_X(x_t = k|x_{\setminus j (t-1)}, x_{j(t-1)} = l_Y) \nonumber \\
- p_X(x_t = k|x_{\setminus j (t-1)}, x_{j(t-1)} = l_X) &= \tilde{\bf X}^j_{k l_Y} \nonumber \\
&\neq -\tilde{\bf Y}^j_{k l_X} \nonumber \\
p_Y(x_t = k|x_{\setminus j(t-1)}, x_{j(t-1)} = l_Y) \nonumber \\ - p_Y(x_t = k|x_{\setminus j (t-1)}, x_{j(t-1)} = l_X) &= \nonumber
\end{align} 
showing the MTD distributions parametrized by ${\bf X}$ and ${\bf Y}$ are not the same, leading to a contradiction so that $\tilde{\bf X} = \tilde{\bf Y}$. The same argument shows that the reduction is unique.

\paragraph{Proof of Proposition \ref{Convex}}
For any two MTD factorizations ${\bf Z}$ and $\tilde{\bf Z}$ and any $x_{kt}$ and $x_{(t-1)}$
\begin{align}
&\sum_{j = 1}^p \left(\alpha \Zj_{x_{kt} x_{j(t-1)}} + (1 - \alpha)  \tilde{\Zj}_{x_{kt} x_{j(t-1)}} \right) \nonumber \\ 
&=  \alpha \sum_{j = 1}^p \Zj_{x_{kt} x_{j(t-1)}} + (1 - \alpha) \sum_{i = 1}^p \tilde{\Zj}_{x_{kt} x_{j(t-1)}} \nonumber \\
&= \alpha p(x_{kt}|x_{(t-1)}) + (1 - \alpha) p(x_{kt}|x_{(t-1)}) \nonumber\\
&= p(x_{kt}|x_{(t-1)}).
\end{align}

\paragraph{Proof of Theorem \ref{optid}}
First, we note that a solution always exists since the log likelihood $L({\bf Z}) = - \sum_{t = 1}^{T}\log \left(z_{x_{jt}} + \sum_{i = 1}^{p} \Zj_{x_{jt} \,\, x_{i(t-1)}} \right) $ and penalty are both bounded below by zero and the feasible set is closed and bounded.
Suppose an optimal solution is ${\bf Z}$ such that there exists some $i$ such that one row, call it $k$, of $\Zj$ does not have a zero element. Let $\alpha = \min(\Zj_{k:})$ be the minimum value in row $k$ and let $\tilde{\Zj}$ be equal to $\Zj$ $\forall i$ except that $\tilde{\Zj}_{k:} = \Zj_{k:} - \alpha$ and $\tilde{z}^j_k = z^{j}_k + \alpha$. Due to the nonidentifiability of the MTD model $L(\tilde{\bf Z}) = L({\bf Z})$, while we have that $\Omega(\tilde{\Zj}) <  \Omega(\Zj)$, implying for $\lambda > 0$
\begin{align}
L(\tilde{\bf Z}) + \lambda \Omega(\tilde{\bf Z}) < L({\bf Z}) + \lambda \Omega({\bf Z}),
\end{align}
showing that ${\bf Z}$ cannot be an optima. 


\end{document}



\maketitle
\section{Proofs}
\begin{definition}
\end{definition}

\begin{proposition}
In both LTD and MTD models, $x_j$ is Granger non-causal for $x_i$ iff the columns of $\Zj$ are all equal.
\end{proposition}
(\emph{proof}) If the columns of $\Zj$ are all equal then for all fixed values of $x_{(t-1) \backslash j}$ the conditional distribution is the same for all values of $x_{(t-1)j}$. If one column is different then the conditional distribution for all values of $x_{(t-1)\backslash j}$ will depend on $x_{(t-1)j}$.
\begin{ntheorem} \label{mtd_ident}
Every MTD distribution has a unique parameterization such that the minimal element in each row of $\Pj$ ($\Zj$) is zero for all $j$.
\end{ntheorem}
(\emph{proof}) 

\begin{proposition}
Every mLTD has a unique parameterization such that first column and last row $\Zj$ are zero for all $j$ and the last element of $\zzero$ is zero.
\end{proposition}
(\emph{proof}) 

\begin{ntheorem}
For any $p$, if $x_i$ is not Granger causal for $x_j$ under $p$, then $x_i$ is not Granger causal for $x_j$ under $p_{Z^{*}}$ for both MTD and mLTD models. 
\end{ntheorem}
(\emph{proof}) Let the optimal solution for $Z^{jk} \,\, \forall k \neq i$ be given by $Z^{jk*}$. For the MTD model let the constraint set be $H$, the optimal $Z^{ji*}$ is given by: 
 \begin{align}
 =&\text{argmax}_{Z^{ji}|\{Z\}} \sum_{x_{jt},x_{t-1}} p(x_{tj} | x_{t-1}) \log p_{\{Z^{jk*}\}_{k \neq i},Z^{ji}}(x_{tj}|x_{t-1}) \\
=& \text{argmax}_{Z^{ji}} \sum_{m = 1}^{|\mathcal{X}_i|} \sum_{x_{jt},x_{(t-1) \backslash i}} p(x_{tj} | x_{t-1}) \log p_{\{Z^{jk*}\}_{k \neq i},Z^{ji}_{:m}}(x_{tj}|x_{(t-1)\backslash i}, x_{(t-1)i} = m) \\
=& \text{argmax}_{Z^{ji}} \sum_{m = 1}^{|\mathcal{X}_i|} \sum_{x_{jt},x_{(t-1) \backslash i}} p(x_{tj} | x_{(t-1)\backslash i}) \log p_{\{Z^{jk*}\}_{k \neq i},Z^{ji}_{:m}}(x_{tj}|x_{(t-1)\backslash i}, x_{(t-1)i} = m)
 \end{align}

\begin{proposition} \label{Convex} 
The set of MTD parameters, ${\bf Z}$, that factorize a conditional distribution $p(x_{tj} | x_{(t-1)})$ is a convex set.
\end{proposition}
(\emph{proof}) For any two MTD factorizations $\{\{\tilde{Z}^{ji}\}_{i = 1}^{p}$ and $\{Z^{ji}\}_{i = 1}^{p}$ and any $x_{tj}$ and $x_{(t-1)}$
\begin{align}
\sum_{i = 1}^p \left(\alpha Z^{ij}_{x_{tj} x_{(t-1)i}} + (1 - \alpha)  \tilde{Z}^{ij}_{x_{tj} x_{(t-1)i}} \right) &= \alpha \sum_{i = 1}^p Z^{ij}_{x_{tj} x_{(t-1)i}} + (1 - \alpha) \sum_{i = 1}^p \tilde{Z}^{ij}_{x_{tj} x_{(t-1)i}} \\
&= \alpha p(x_{tj}|x_{(t-1)}) + (1 - \alpha) p(x_{tj}|x_{(t-1)}) \\
&= p(x_{tj}|x_{(t-1)}).
\end{align}

\begin{ntheorem} 
For any $\lambda > 0$ and $\Omega(Z)$ that does not depend on $\zzero$ and is increasing with respect to the absolute value of entries in $\Zj$, the solution to the problem in Eq. \refeq{ident_mtd} is contained in the set of identifiable MTD models described in Theorem \ref{mtd_ident}. 
\end{ntheorem}

(\emph{proof}) First, we note that a solution always exists since the log likelihood $L(Z) = - \sum_{t = 1}^{T}\log \left(z^j_{x_{tj}} + \sum_{i = 1}^{p} Z^{ji}_{x_{tj} \,\, x_{(t-1)i}} \right) $ and penalty are both bounded below by zero and the feasible set is closed and bounded.
Suppose an optimal solution is $Z$ such that there exists some $i$ such that one row, call it $k$, of $Z^{ji}$ does not have a zero element. Let $\alpha = \min(Z^{ji}_{k:})$ be the minimum value in row $k$ and let $\tilde{Z}^{ji}$ be equal to $Z^{ji}$ $\forall i$ except that $\tilde{Z}^{ji}_{k:} = Z^{ji}_{k:} - \alpha$ and $\tilde{z}^j_k = z^{j}_k + \alpha$. Due to the nonidentifiability of the MTD model $L(\tilde{Z}) = L(Z)$, while we have that $||\tilde{Z}^{ji}||_{2} <  ||Z^{ji}||_2$, implying for $\lambda > 0$
\begin{align}
L(\tilde{Z}) + \lambda \Omega(\tilde{Z}) < L(Z) + \lambda \Omega(Z),
\end{align}
showing that $Z$ cannot be an optima. 
\subsection{BIC for MTD}
\section{Optimization}
\subsection{Group lasso MTD}
We show that a group lasso over entries in $\Zj$ is a convex relaxation to the $L_0$ norm over $\gamma_{1:p}$.   For simplicity assume $m_j = m \,\,\, \forall j$. Due to the equality and greater than zero constraints
\begin{align}
||\gamma_{1:p}||_0 &= ||\left({\bf 1}^T \text{vec}({\bf Z}^1),\ldots, {\bf 1}^T \text{vec}({\bf Z}^p) \right)||_0 \\
&= \text{rank}(H_1^T H_1) \\
&= \text{rank}(H_1)
\end{align}
where 
\begin{align}
H_1 = \left(\begin{array}{c c c c}
\text{vec}({\bf Z}^1) & 0 & \ldots & 0 \\
0 &  \text{vec}({\bf Z}^2) & \ldots & 0 \\
0 & \ldots & \ddots & \vdots \\
0 & \ldots  & \ldots & \text{vec}({\bf Z}^p)
\end{array} \right)
\end{align}
Thus we can use the nuclear norm on $H_1$ as a convex relaxation, 
\begin{align}
||H_1||_{*} = \sum_{i = 1}^{p} ||\Zj||_F.
\end{align} 
\subsection{Projected gradient MTD}
The gradient of the MTD model over the feasible set is given by:
\begin{align}
\frac{d L}{d Z^j_{x^{'},x^{''}}} = \sum_{t = 1}^{T} 1_{x_{it} = x^{'}, x_{j(t-1)} = x^{''}} \frac{1}{\zzero_{x_{it}} + \sum_{j = 1}^{p} \Zj_{x_{it},x_{j(t-1)}}} + \lambda \frac{d \Omega}{d Z^j_{x^{'},x^{''}}}.
\end{align}
Note that for the $L_1$ and $L_2$ norms $\Omega(Z)$ is not differentiable when elements are equal to zero. However, note that due to our problem constraints we have that $\Zj \geq 0$. Since the point of non-differentiability occurs when elements are identically zero, we modify the problem constraints so that $\Zj \geq \epsilon$ for some small $\epsilon$, so we may ignore the non-differentiability. Following the notation from the main text, let the set $C = \{\tilde{z} | \tilde{z}\geq \epsilon, (I_p \otimes A) \tilde{z} = 0, 1^T\tilde{z} = m\}$. We perform projected gradient descent:
\begin{align}
\tilde{z}^{k + 1} = P_{C}\left(\tilde{z}^{k} - \gamma_k \frac{d L}{d \tilde{z}}\right)
\end{align}
where $\gamma_k$ is the step size, which we chose by line search, and $P_{C}(x)$ is the projection of $x$ onto the set $C$:
\begin{equation*}
\begin{aligned}
& \underset{z}{\text{minimize}} \,\,\,\,\, ||z - x||_2^2 \\
& \text{subject to} \,\,\,\,\, z \geq \epsilon,\,\,\,\,\, (I_p \otimes A)z  = 0,\,\,\,\, 1^T z = m.
\end{aligned}
\end{equation*}
This is a quadratic program which we solve using the dual method of Goldfarb and Idnani (1982, 1983) as implemented in the R quadratic programming package \emph{quadprog}. Note that simply projecting onto the simplex may be done efficiently in $ (\mathcal{O}) n \log n$ time \cite{Duchi}. Perhaps there is a similar type algorithm for fast projection onto $C$. We also utilize a simple acceleration method which we find vastly improves convergence in practice.
\subsection{Projected gradient mLTD}